\documentclass[hidelinks,onefignum,onetabnum]{siamart250211}

\usepackage{hyperref}       
\usepackage{url}            
\usepackage{booktabs}       
\usepackage{graphicx}
\usepackage{doi}
\usepackage{subcaption}

\usepackage{amsmath, amsfonts, bbm, bm}
\usepackage{graphicx}

\usepackage{lipsum}
\usepackage{amsfonts}
\usepackage{graphicx}
\usepackage{epstopdf}
\usepackage{algorithmic}
\ifpdf
  \DeclareGraphicsExtensions{.eps,.pdf,.png,.jpg}
\else
  \DeclareGraphicsExtensions{.eps}
\fi

\newsiamremark{remark}{Remark}
\newsiamremark{hypothesis}{Hypothesis}
\crefname{hypothesis}{Hypothesis}{Hypotheses}
\newsiamthm{claim}{Claim}
\newsiamremark{fact}{Fact}
\crefname{fact}{Fact}{Facts}

\headers{Exploiting Sparse Conditional Dependence in Remote Sensing}{K. M. Leung, J. Susiluoto, R. Eckert, Y. Marzouk}

\title{Exploiting Sparse Conditional Dependence in Remote Sensing Observing Systems}

\author{Kelvin M. Leung\thanks{Massachusetts Institute of Technology
  (\email{kmleung@mit.edu}).}
\and Jouni Susiluoto\thanks{NASA Jet Propulsion Laboratory}
\and Regina Eckert\thanks{NASA Jet Propulsion Laboratory}
\and Youssef Marzouk\thanks{Massachusetts Institute of Technology}}

\usepackage{amsopn}

\usepackage{etoolbox}
\AtBeginDocument{
  \renewcommand{\url}[1]{}  
}

\usepackage{xcolor} 
\usepackage[textsize=tiny]{todonotes}

\ifpdf
\hypersetup{
  pdftitle={An Example Article},
  pdfauthor={K. M. Leung, J. Susiluoto, R. Eckert, and Y. Marzouk}
}
\fi

\begin{document}

\maketitle

\begin{abstract}
We develop a novel filtering algorithm for remote sensing inverse problems that exploits sparsity in the precision (inverse covariance) matrix arising from conditional independence structures in spatio-temporal satellite data. Motivated by NASA’s Orbiting Carbon Observatory-2 (OCO-2), Earth Surface Mineral Dust Source Investigation (EMIT), and upcoming Explorer for Artemis Geology Lunar and Earth (EAGLE) missions, we formulate a Bayesian inference framework for retrieving atmospheric and surface parameters from hyperspectral radiance measurements. 
Unlike traditional methods that operate on covariances and ignore spatio-temporal dependencies, our approach models the full spatio-temporal state using sparse Gaussian graphical models, which corresponds to a block-sparse precision matrix.
We derive an information filter based on this structure, enabling scalable joint inference over large spatial grids with uncertainty propagation. 
We demonstrate the effectiveness of the algorithm in two distinct remote sensing contexts. First, we apply the method to synthetic data configured to mimic the OCO-2 mission, focusing on retrievals of atmospheric CO$_2$, and observe a 50\% reduction in CO$_2$ retrieval error compared to single-pixel methods.
Then, we evaluate the algorithm using imaging spectroscopy data from the Surface Biology and Geology (SBG) High Frequency Time Series (SHIFT) campaign to retrieve surface reflectance, achieving a 20\% reduction in retrieval error relative to baseline approaches.
\end{abstract}



\section{Introduction}

Every year, an increasing number of satellites dedicated to planetary and environmental monitoring are launched into space.
Orbiting Carbon Observatory-2 (OCO-2) is a NASA mission launched in 2014 that uses hyperspectral imagery to monitor atmospheric carbon dioxide concentrations \cite{eldering_orbiting_nodate}. 
Earth Surface Mineral Dust Source Investigation (EMIT) is an ongoing NASA mission to map the mineral composition in arid dust regions using imaging spectroscopy \cite{thompson_-orbit_2024}.
Explorer for Artemis Geology Lunar and Earth (EAGLE) is an upcoming mission aimed at monitoring Earth surface properties including ecosystem health, snow and ice melt, and mineral content under NASA's Earth System Observatory (ESO) Surface Biology and Geology (SBG) focus area \cite{cawse-nicholson_nasas_2021, cawse-nicholson_surface_2023}. 

Satellite measurements consist of radiances across a spectrum, collected by an imaging spectrometer.
The quantities of interest – parameters describing atmospheric conditions and surface reflectance – are connected to observations through a physics-based radiative transfer forward model. 
Retrieving these parameters requires solving a Bayesian inverse problem, essentially performing atmospheric correction. 
State-of-the-art methods use single-pixel optimal estimation (OE) \cite{thompson_optimal_2018}, which is computationally efficient but ignores spatial and temporal correlations.



This setup naturally frames as a spatio-temporal filtering or smoothing problem. The underlying state consists of correlated surface and atmospheric variables that are unobservable directly. 
Instead, a satellite periodically revisits locations, allowing iterative state estimation via inverse problems. While extensions of OE to spatial fields exist \cite{hobbs_spatial_2021, eckert_spatially_2023, zilber_spatial_2022}, little work has addressed full spatio-temporal fields. 

We close this gap with a scalable Bayesian inference framework for remote sensing systems that collect spatial imagery at regular intervals. Our method models the spatio-temporal state using sparse Gaussian graphical models, which naturally yield an efficient filtering algorithm enabling uncertainty-aware inference across space and time. We demonstrate its effectiveness on two applications aligned with the OCO-2, EMIT, and EAGLE missions.

Sequential estimation in large-scale spatio-temporal systems is well-studied across climate science, geophysics, and remote sensing. Filtering approaches, such as the Kalman filter and its nonlinear extensions, sequentially estimate an evolving state by incorporating observations \cite{houtekamer_data_1998, evensen_ensemble_2003, sarkka_bayesian_2013}, performing at each step a forecast via model dynamics and a conditioning on observations. However, even under a Gaussian assumption, these computations become prohibitive in high dimensions due to the cost of inverting and propagating large covariance matrices.

A common approach to addressing this challenge is to reduce the effective spatial dimension at each time step. Several strategies have been developed for this purpose.
The first is low-rank modeling, where the spatial field is represented by a fixed number of deterministic basis functions multiplied by random coefficients, the canonical example being fixed rank kriging \cite{banerjee_gaussian_2008, finley_improving_2009, wikle_spatio-temporal_2019, cressie_fixed_2008}.
A second strategy approximates the covariance as a sparse matrix to reduce the cost of matrix inversions, including spatial partitioning \cite{heaton_case_2019}, covariance tapering \cite{kaufman_covariance_2008, furrer_covariance_2006}, and inducing point methods \cite{wilson_kernel_2015, rasmussen_gaussian_2006}. Sparse approximations to the precision matrix have also been studied, including multiresolution approximations \cite{katzfuss_multi-resolution_2017, nychka_multiresolution_2015}, Vecchia approximations \cite{vecchia_estimation_1988, stein_approximating_2004, katzfuss_general_2021}, nearest-neighbor Gaussian processes \cite{datta_nearest-neighbor_2016}, and the SPDE approach \cite{lindgren_explicit_2011, lindgren_spde_2022}.

These techniques can be embedded within data assimilation frameworks. In ensemble-based filtering, localization tapers the background covariance or Kalman gain to zero beyond a spatial radius, assuming distant state variables are conditionally independent given observations \cite{gaspari_construction_1999, houtekamer_data_1998, ott_local_2004, bishop_adaptive_2001, vishny_high-dimensional_2024}. This is standard in operational weather forecasting, where ensemble size is much smaller than state dimension \cite{bannister_review_2017, hamill_distance-dependent_2001}. Adaptive variants adjust the localization scale based on flow-dependent structure or sampling noise \cite{anderson_localization_2012, bishop_ensemble_2009, bishop_ensemble_2009-1, bishop_adaptive_2011}, and the approach can be viewed as a filtering analog to covariance tapering \cite{furrer_covariance_2006}.

Finally, rather than approximating the covariance, some methods work directly in the information domain by maintaining the precision matrix. This idea appears in early filtering literature \cite{anderson_optimal_1982, jazwinski_stochastic_1970} and is especially useful when conditional independencies yield sparse precisions. Recent developments such as the inverse EnKF \cite{singh_inverse_2023, fang_inverse_2025, lunde_ensemble_2025} show how ensemble methods can be efficiently formulated in this information form.

\subsection{Contributions}

We introduce a scalable Bayesian framework for spatio-temporal inference in remote sensing, grounded in physical modeling of spatial and temporal dependencies. 
By deriving the conditional dependence structure of the state variables \textit{a priori}, we induce sparsity in the precision matrix using Gaussian graphical models. Our primary contributions are:
\begin{enumerate}
\item We develop an efficient filtering framework for remote sensing observing systems that exploits sparsity induced by knowledge of the physical model to enable scalable inference over large spatio-temporal grids, integrating graphical models and information filtering into a flexible framework for Earth observation systems.
\item We validate the framework on realistic Earth science applications drawn from NASA's OCO-2, EMIT, and EAGLE missions, demonstrating both scalability and improved retrieval accuracy over existing methods.
\end{enumerate}

Most closely related to our work, \cite{zilber_spatial_2022} introduces spatial surface reflectance retrievals for VSWIR remote sensing using a dense covariance-based atmospheric prior. Our method extends this in two key ways: by deriving a sparse pentadiagonal atmospheric precision matrix from explicit conditional independence assumptions, enabling a 100-fold increase in problem size, from a $3\times 3$ to a $30\times 30$ spatial grid, and by incorporating temporal filtering to enable full spatio-temporal inference.

We present the problem setup and conditional independence assumptions in Section~\ref{sec:setup}, the filtering algorithm in Section~\ref{sec:methodology}, numerical results in Section~\ref{sec:results}, and extensions in Section~\ref{sec:discussion}.

\section{Conditional Independence in Remote Sensing Problems}
\label{sec:setup}


In this section, we identify the conditional independence relationships inherent in remote sensing retrievals and construct a corresponding graphical model to represent the problem structure. 
Probabilistic graphical models characterize dependencies among large sets of variables, and in the Gaussian case, the conditional independence structure is directly reflected in the sparsity pattern of the precision matrix.

\subsection{Probabilistic Graphical Models}
A graph $\mathcal{G} = (\mathcal{V}, \mathcal{E})$ consists of nodes $i\in\mathcal{V}$ connected by edges $\mathcal{E} = \{(i,j) : i, j \in \mathcal{V}\}$. 
Edges may be directed, encoding causal or conditional dependence, or undirected, encoding conditional independence. 
The remote sensing problem involves both causal dependencies (time evolution and forward model structure) and symmetric spatial dependencies, and is therefore naturally represented as a hybrid model.

\textit{Directed} graphical models (DAGs) represent conditional dependence relations among random variables \cite{lauritzen_graphical_1996, jordan_graphical_2004, murphy_machine_2012}.
Each directed edge $(i, j)$ implies that node $j$ is conditionally dependent on node $i$, and if all paths from a node set $A$ to node $j$ pass through $i$, then $j$ is conditionally independent of $A$ given $i$.
The DAG implies a factorization of the joint distribution:
\begin{align}
    \label{eq:dag_joint}
    p(\mathbf{x}_\mathcal{V}) = \prod_{i\in \mathcal{V}} p(x_i | x_{\pi(i)}),
\end{align}
where $\pi(i)$ denotes the parents of node $i$.

\textit{Undirected} graphical models, or Markov random fields (MRFs), represent conditional independence through undirected edges \cite{lauritzen_graphical_1996, wainwright_graphical_2008}, where $(i,j) \in \mathcal{E}$ implies $(j,i) \in \mathcal{E}$.
Sets $A$ and $B$ are conditionally independent given $C$ if all paths between them are blocked upon removing $C$.
The joint distribution factorizes over maximal cliques:
\begin{align}
    \label{eq:mrf_joint}
    p(\mathbf{x}_\mathcal{V}) = \frac{1}{Z} \prod_{c \in \text{cl}^*(\mathcal{G})} \psi_c(x_c),
\end{align}
where $\psi_c(\cdot)$ is a potential function over clique $c$ and $Z$ is a normalization constant.
This factorization is particularly useful for Gaussian models, where the zero pattern of the precision matrix corresponds exactly to the graph structure.

In our spatio-temporal remote sensing setting, temporal and forward model relationships are encoded with directed edges, while spatial dependencies among atmospheric variables are encoded with undirected edges. 
This hybrid structure captures both the causal flow of time and symmetric spatial dependencies, enabling efficient and scalable inference.


\subsection{Remote Sensing Retrievals}

We begin with a detailed overview of the remote sensing retrieval problem before constructing the associated spatio-temporal graph. 
The retrieval task is an inverse problem: for each pixel in a hyperspectral image, we infer the surface and atmospheric parameters that explain the observed radiance.

Let $\mathbf{y} \in \mathbbm{R}^{n_y}$ denote radiances (in $\mu \text{W}\text{nm}^{-1} \text{sr}^{-1}\text{cm}^{-1}$) recorded by the imaging spectrometer.
These radiances are related to the physical state $\mathbf{x}$, consisting of surface and atmospheric parameters, through a physics-based radiative transfer forward model $\mathbf{y} = f(\mathbf{x}) + \epsilon$, where  $\epsilon$ is additive observation noise. 

\begin{figure}
    \centering
    \includegraphics[width=\textwidth]{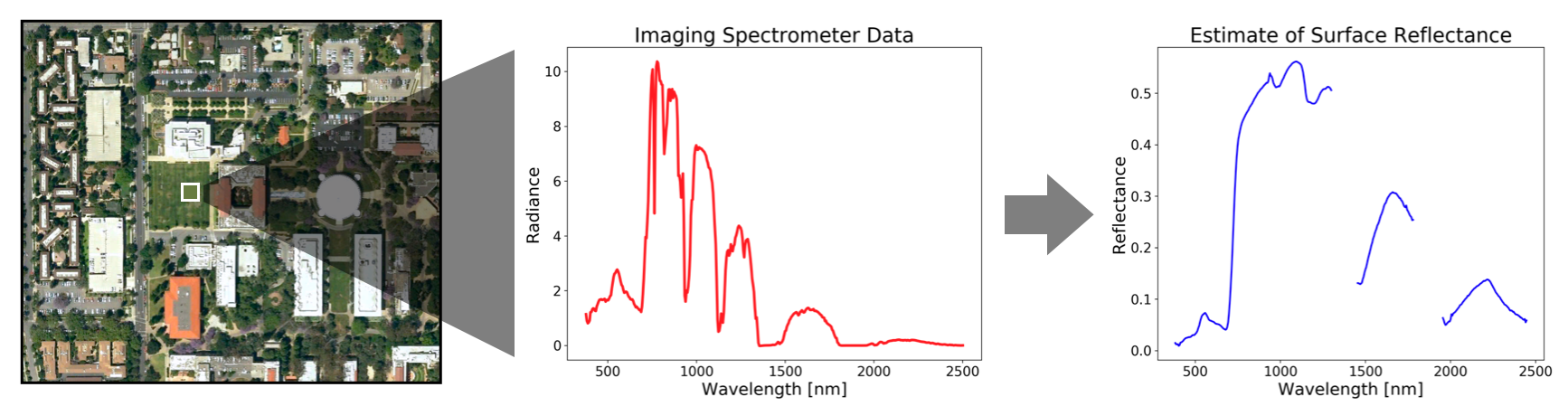}
    \caption{Retrieval of surface reflectance for a single pixel.}
    \label{fig:retrievalProcess}
\end{figure}

Our spatio-temporal retrieval method extends single-pixel optimal estimation (OE) \cite{thompson_optimal_2018}. 
While retrievals include both surface and atmospheric components, specific missions may prioritize one (e.g., OCO-2 prioritizes atmosphere, EMIT and EAGLE prioritize surface).
Figure~\ref{fig:retrievalProcess} illustrates single-pixel surface reflectance retrieval, where the posterior, $p(\mathbf{x} | \mathbf{y}) \propto p(\mathbf{y} | \mathbf{x} )\, p(\mathbf{x})$,
is approximated via a MAP estimate with a Laplace approximation for the posterior covariance.

The state vector $\mathbf{x} = [\mathbf{s}, \mathbf{a}]^\top \in \mathbbm{R}^n$ concatenates surface parameters $\mathbf{s} \in \mathbb{R}^{n_s}$ and atmospheric parameters $\mathbf{a} \in \mathbb{R}^{n_a}$, with spatial indices $j = 1, \dots, m$ as subscripts and temporal indices $t = 1, \dots, T$ as superscripts.
The full spatial state at time $t$ is:
\begin{align}
    \mathbf{x}_{1:m}^t = [\,\mathbf{s}_1, \mathbf{a}_1 \,\, | \,\, \mathbf{s}_2, \mathbf{a}_2 \,\, | \,\, \dots \,\, | \,\,\mathbf{s}_m, \mathbf{a}_m ]^\top
\end{align}

Both $\mathbf{s}$ and $\mathbf{a}$ are assigned Gaussian priors: surface parameters follow a joint multivariate Gaussian capturing cross-component correlations, while atmospheric parameters are modeled as independent Gaussians.


The forward model $f(\cdot) = [f_1(\cdot), \dots, f_{n_y}(\cdot)]^\top$ maps state parameters to expected radiances, capturing light propagation through the atmosphere to the spectrometer.
Observation noise $\epsilon$ is modeled as zero-mean Gaussian with diagonal covariance $\Gamma_\epsilon$, typically determined by instrument signal-to-noise ratios.


\subsection{Conditional Independence Assumptions}

Extending single-pixel retrievals to the spatio-temporal setting requires structural independence assumptions that enable problem decomposition and computational savings while still exploiting spatial and temporal correlations.

For atmospheric parameters, we assume temporal independence across satellite overpasses, motivated by multi-day revisit intervals during which atmospheric conditions decorrelate due to weather dynamics. Spatial correlations within a single time step are retained.



For surface parameters, we assume spatial independence across pixels, reflecting the heterogeneity of land cover – particularly in vegetated regions, which are a primary focus of missions like EAGLE. Surface reflectance does evolve smoothly over time due to seasonal dynamics, so temporal correlations are retained.


These assumptions reflect common characteristics in Earth observation data and generalize to a broad class of joint atmospheric-surface retrieval problems from missions with multi-day revisit cycles, such as OCO-2, EMIT, and EAGLE. We now formalize this conditional independence structure as a probabilistic graphical model.


\begin{figure}
    \centering
    \begin{minipage}[c]{0.65\linewidth}
        \begin{subfigure}{\linewidth}
            \centering
            \includegraphics[width=\linewidth]{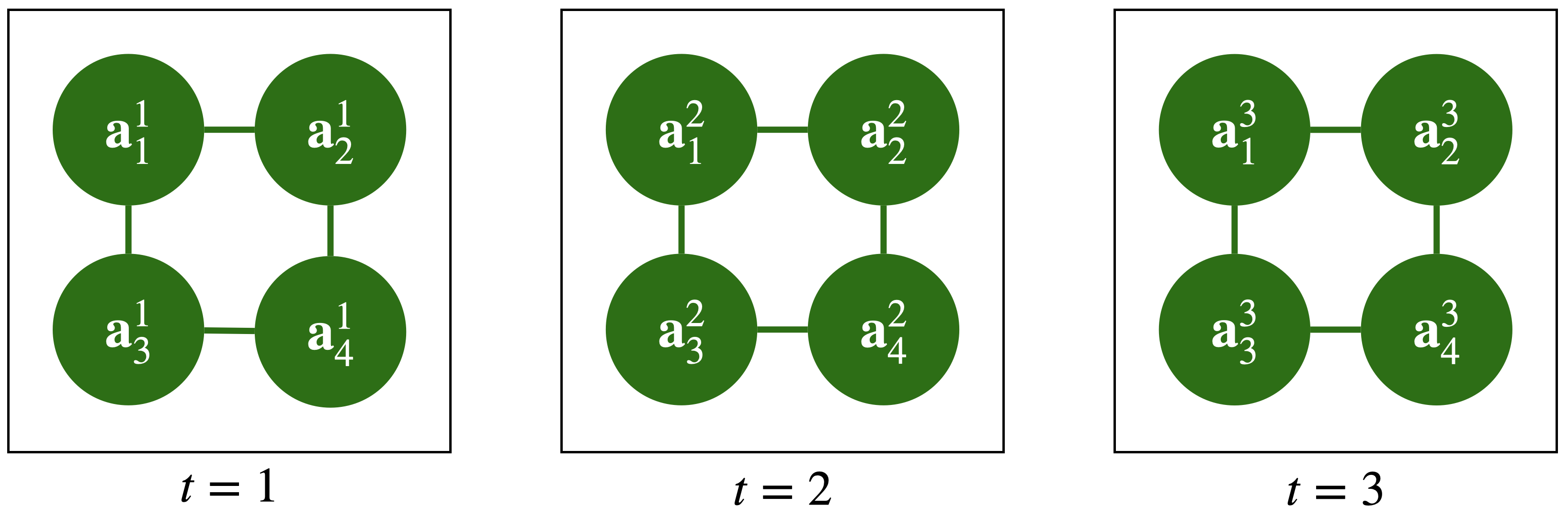}
            \caption{Undirected graphical structure of atmospheric parameters over $m=4$ spatial locations and $T=3$ time steps.}
            \label{fig:graph_atm}
        \end{subfigure}
        \vspace{1em}
        \begin{subfigure}{\linewidth}
            \centering
            \includegraphics[width=\linewidth]{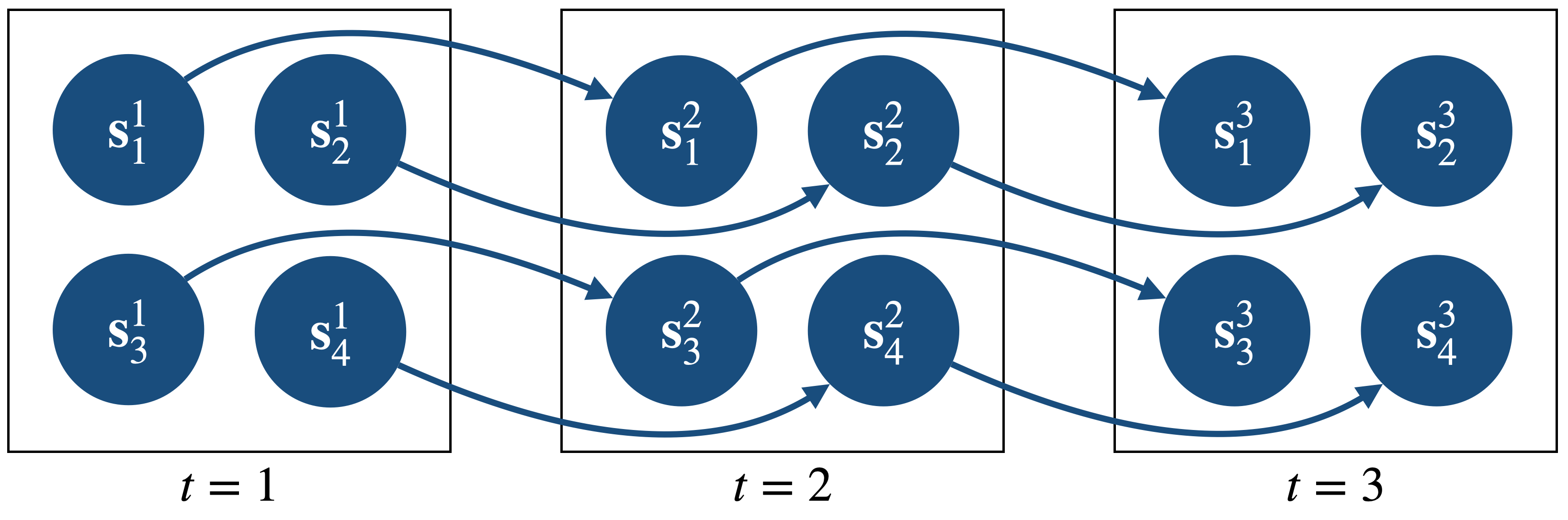}
            \caption{Directed graphical structure of surface parameters over $m=4$ spatial locations and $T=3$ time steps.}
            \label{fig:graph_refl}
        \end{subfigure}
    \end{minipage}
    \hfill
    \begin{minipage}[c]{0.3\linewidth}
        \begin{subfigure}{\linewidth}
            \centering
            \includegraphics[width=0.4\linewidth]{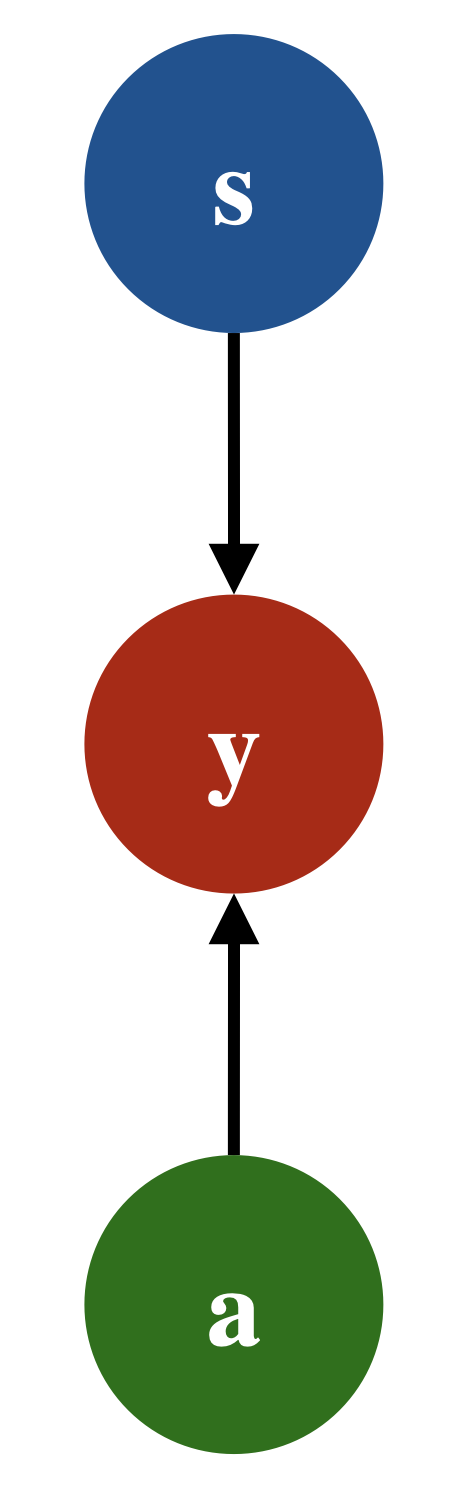}
            \caption{Dependencies among surface, atmosphere, and observed radiance.}
            \label{fig:graph_yas_vertical}
        \end{subfigure}
    \end{minipage}
    \caption{Graphical model structure for the spatio-temporal remote sensing problem.}
    \label{fig:graphs_combined}
\end{figure}

Figure~\ref{fig:graph_atm} illustrates the graphical structure of the atmospheric parameters over $m=4$ spatial locations and $T=3$ time steps.
We assign a grid structure to the atmosphere within each time step, so each variable is conditionally independent of all others given its four spatial neighbors. Edges are undirected, reflecting symmetric spatial relationships, and no temporal edges are included.


Figure~\ref{fig:graph_refl} shows the surface graph for the same dimensions. Here, we introduce directed edges that connect surface parameters at the same spatial location across successive time steps, encoding temporal dependence. No spatial edges are included, consistent with the spatial independence assumption.




The inter-variable dependencies at a single pixel are shown in Figure~\ref{fig:graph_yas_vertical}. 
As given in Equation~\eqref{eq:fwd}, the observation $\mathbf{y}$ depends on both surface and atmospheric parameters through $f(\cdot)$, encoded as directed edges from $\mathbf{s}$ and $\mathbf{a}$ to $\mathbf{y}$, giving the factorization $p(\mathbf{s}, \mathbf{a},\mathbf{y}) = p(\mathbf{y} | \mathbf{s}, \mathbf{a})\, p(\mathbf{s}) \, p(\mathbf{a})$.


\subsection{Filtering Distribution}

\begin{figure}
    \centering
    \includegraphics[width=0.8\textwidth]{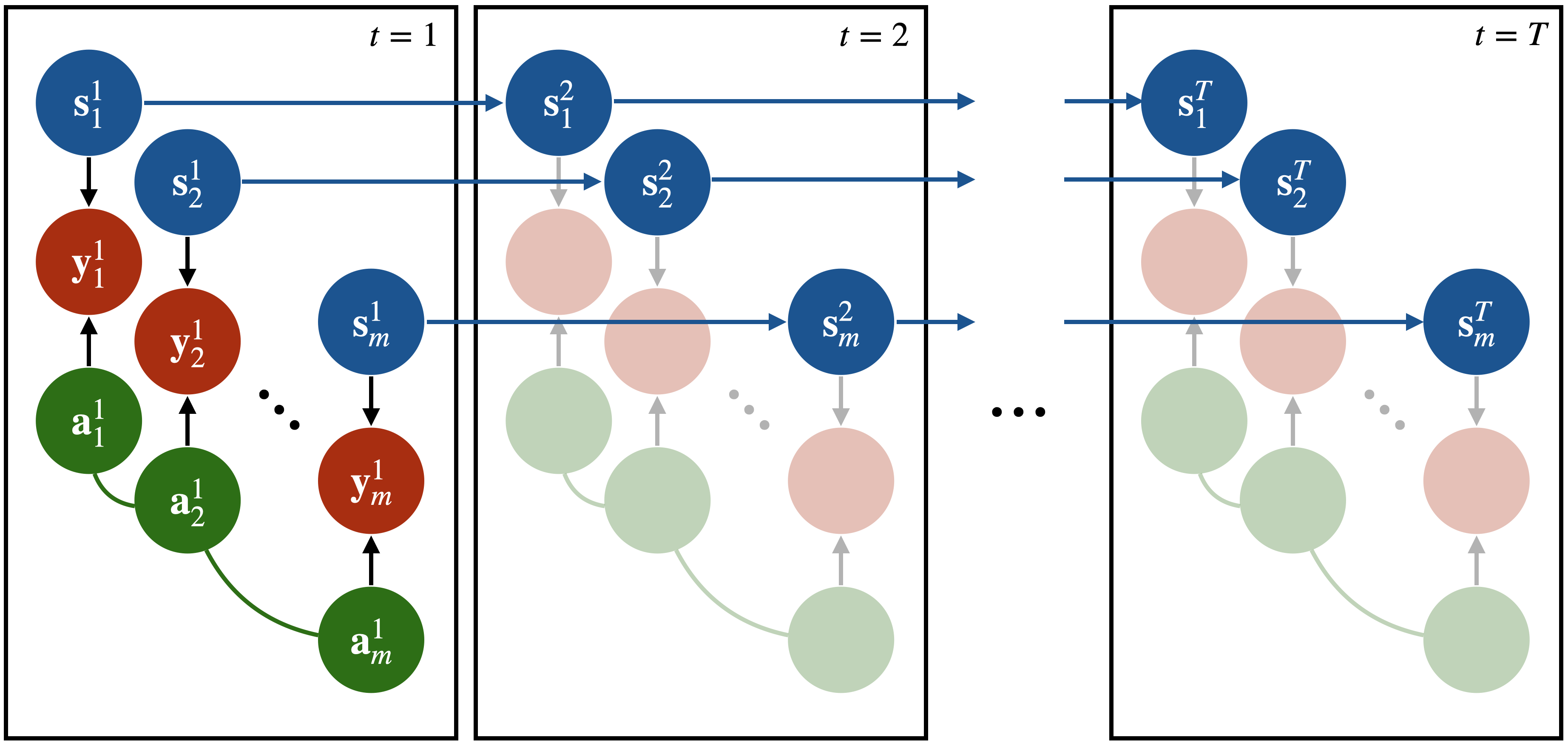}
    \caption{Graphical model of surface and atmospheric parameters and observations across space and time.}
    \label{fig:graph_full}
\end{figure}

The full undirected graphical model over $m$ spatial locations and $T$ time steps is shown in Figure~\ref{fig:graph_full}, with atmospheric parameter independence omitted for clarity.
 The structure resembles a dynamical system, where surface states $\mathbf{s}_j$ evolve over time and are coupled through observations $\mathbf{y}_j$ and atmospheric states $\mathbf{a}_j$.

The system can be interpreted as a spatio-temporal dynamical model with latent states $\mathbf{x}_{1:m} = [\mathbf{s}_{1:m},\mathbf{a}_{1:m}]^\top$.
The filtering goal is to sequentially estimate:
\begin{align}
    p(\mathbf{x}_{1:m}^t | \mathbf{y}_{1:m}^{1:t}) = p(\mathbf{s}_{1:m}^t, \mathbf{a}_{1:m}^t  | \mathbf{y}_{1:m}^{1:t}).
\end{align}

The conditional independence structure encoded by the graph allows this distribution to be factored using the DAG and MRF rules from Equations~\eqref{eq:dag_joint} and~\eqref{eq:mrf_joint} (full derivation in Appendix~\ref{sec:filterderivation}):
\begin{align}
    \label{eq:filter}
    p(\mathbf{x}_{1:m}^t | \mathbf{y}_{1:m}^{1:t}) \propto p(\mathbf{y}_{1:m}^t | \mathbf{s}_{1:m}^t, \mathbf{a}_{1:m}^t) \, p(\mathbf{s}_{1:m}^t | \mathbf{y}_{1:m}^{1:t-1}) \, p(\mathbf{a}_{1:m}^t),
\end{align}
with the components:
\begin{align}
    p(\mathbf{y}_{1:m}^t | \mathbf{s}_{1:m}^t, \mathbf{a}_{1:m}^t) &= \prod_{j=1}^m p(\mathbf{y}_j^t | \mathbf{s}_j^t, \mathbf{a}_j^t), \\
    p(\mathbf{s}_{1:m}^t | \mathbf{y}_{1:m}^{1:t-1}) &= \int_{\mathbf{s}^{t-1}_{1:m}} \left( \prod_{j=1}^m p(\mathbf{s}_j^t | \mathbf{s}_j^{t-1}) \right) \,p(\mathbf{s}_{1:m}^{t-1} | \mathbf{y}_{1:m}^{1:t-1}) \, d\mathbf{s}_{1:m}^{t-1}
\end{align}

Each term corresponds to a distinct aspect of the filtering procedure. The likelihood $p(\mathbf{y}_{1:m}^t | \mathbf{s}_{1:m}^t, \mathbf{a}_{1:m}^t)$ factorizes across spatial locations due to conditional independence of observations given the local state. The forecast term $p(\mathbf{s}_{1:m}^t | \mathbf{y}_{1:m}^{1:t-1})$ propagates the previous posterior through the temporal model, integrating out the previous surface state.
The atmospheric prior $p(\mathbf{a}_{1:m}^t)$ models each spatial field $a_{i,1:m}$ as a multivariate Gaussin encoding spatial correlation, forming a single clique in the graph. 


Together, these terms allow the algorithm to update each pixel's surface state using information from prior time steps while enforcing spatial regularity in the atmosphere. We derive a filtering algorithm operating in the information (precision) domain, tailored to the sparsity structure of remote sensing retrieval problems.


\section{Information Filter for Imaging Spectroscopy}
\label{sec:methodology}

We represent the problem as a dynamical system and demonstrate how an information filter \cite{anderson_optimal_1982, jazwinski_stochastic_1970, lunde_ensemble_2025} can be effectively applied by leveraging the sparsity in the precision matrix induced by conditional independence.

\subsection{Dynamical Model}

Unlike many nonlinear filtering setups, our system features linear state dynamics paired with a nonlinear observation model.
This spatial observation function $F: \mathbbm{R}^{nm} \rightarrow \mathbbm{R}^{n_y m}$ applies the forward model $f$ independently to each pixel:
\begin{align}
    F(\mathbf{x}_{1:m}) = \left[ f(\mathbf{x}_1)^\top \dots f(\mathbf{x}_m)^\top \right]^\top.
\end{align}
The full dynamical model is:
\begin{align}
    \label{eq:dynsystem}
    \mathbf{y}_{1:m}^t = F(\mathbf{x}_{1:m}^t) + \varepsilon^t \nonumber\\
    \mathbf{x}_{1:m}^{t+1} = M^t \mathbf{x}_{1:m}^t + \xi^t,
\end{align}
where $\varepsilon^t \sim \mathcal{N}(0, \Sigma_\epsilon)$ is measurement noise and $\xi^t \sim \mathcal{N}(\mu_\xi^t, \Sigma_\xi^t)$ is state dynamics uncertainty.
$M^t$, $\mu_\xi^t$, $\Sigma_\xi^t$ are chosen based on problem-specific knowledge.

From the graphical model in Section~\ref{sec:setup}, atmospheric parameters are assumed temporally independent and thus do not have a time-based filter update. 
In contrast, surface parameters evolve over time but lack known physical dynamics, motivating a statistical model for their temporal evolution. We define the state transition model using two tunable hyperparameters $\alpha$ and $\beta$:
\begin{align}
    M^t = 
    \begin{bmatrix}
        \alpha I_{n_s} & \\
        & \mathbf{0}
    \end{bmatrix},
    \hspace{20pt}
    \mu_\xi^t=
    \begin{bmatrix}
        (1-\alpha) \mu_s \\
        \mu_a
    \end{bmatrix},
    \hspace{20pt}
    \Sigma_\xi^t=
    \begin{bmatrix}
        (1-\alpha)^2 \beta^2 \Sigma_s & \\
        & \Sigma_a
    \end{bmatrix},
\end{align}
where $I_{n_s}$ denotes the identity matrix of size $n_s$.

This model ensures that the posterior at time $t$, denoted $t|t$, informs the prior for the next step $t+1|t$. 
However, directly using the posterior can yield an overconfident prior – surface properties may vary smoothly over time but can also change abruptly due to weather events or human activity. To prevent underestimating uncertainty in such cases, we define the prior at $t+1$ as a convex combination of the posterior at $t$ and the initial prior, modulated by the empirically tuned hyperparameters $\alpha$ and $\beta$.

The resulting updates to the mean and covariance are: 
\begin{align}
    \label{eq:meanforecast}
    \mu^{t+1|t} &= 
    \begin{bmatrix}
        \alpha I_{n_s} & \\
        & \mathbf{0}
    \end{bmatrix}
    \begin{bmatrix}
        \mu_s^{t|t} \\
        \mu_a^{t|t}
    \end{bmatrix} + 
    \begin{bmatrix}
        (1-\alpha) I_{n_s} & \\
        & I_{n_a}
    \end{bmatrix}
    \begin{bmatrix}
        \mu_s\\
        \mu_a
    \end{bmatrix} \\
    \label{eq:covforecast}
    \Sigma^{t+1|t} &= 
    \begin{bmatrix}
        \alpha^2 I_{n_s} & \\
        & \mathbf{0}
    \end{bmatrix}
    \begin{bmatrix}
        \Sigma_s^{t|t} & \\
        & \Sigma_a^{t|t}
    \end{bmatrix} + 
    \begin{bmatrix}
        (1-\alpha)^2 \beta^2 I_{n_s} & \\
        & I_{n_a}
    \end{bmatrix}
    \begin{bmatrix}
        \Sigma_s & \\
        & \Sigma_a
    \end{bmatrix} 
\end{align}

We perform sequential updates of the state $\mathbf{x} = [\mathbf{s}, \mathbf{a}]^\top$ using a filtering algorithm composed of two steps per time point: (i) inference, where the state is updated given observations, and (ii) forecasting, where the state is propagated using the model dynamics. 
Our method operates in the precision domain rather than the covariance domain, exploiting the sparsity induced by the conditional independence structure in Section~\ref{sec:setup}.

\subsection{Sparsity of the prior precision}


For jointly Gaussian variables, conditional independence implies sparsity in the precision matrix. Working in the information form of the filter exploits this sparsity, reducing both computational complexity and memory usage.


We illustrate the structure of $\pmb\Lambda^1$, the prior precision at $t=1$, and show in subsequent subsections that this structure is preserved through both filter steps.

The precision of the full state over space, $\pmb\Lambda \in \mathbbm{R}^{nm \times nm}$, is arranged with inner index $i = 1,\dots, n$ over state components and outer index $j= 1,\dots, m$ over spatial locations, yielding $m \times m$ spatial blocks of size $n \times n$

At $t=1$, surface and atmospheric priors are assumed independent and their precisions are constructed separately.
We define the spatial prior covariance for the surface parameters as $\Sigma_s \in \mathbbm{R}^{mn_s \times mn_s}$.
Since surface parameters are spatially independence, the prior covariance $\Sigma_s = \text{diag}\left( \Gamma_{s_1}, \dots, \Gamma_{s_m} \right)$ is block diagonal, with block-diagonal precision $\Sigma_s^{-1} = \text{diag}\left( \Gamma_{s_1}^{-1}, \dots ,\Gamma_{s_m}^{-1} \right)$.

\begin{figure}
    \centering
    \begin{subfigure}[b]{0.49\linewidth}
        \centering
        \includegraphics[width=0.5\linewidth]{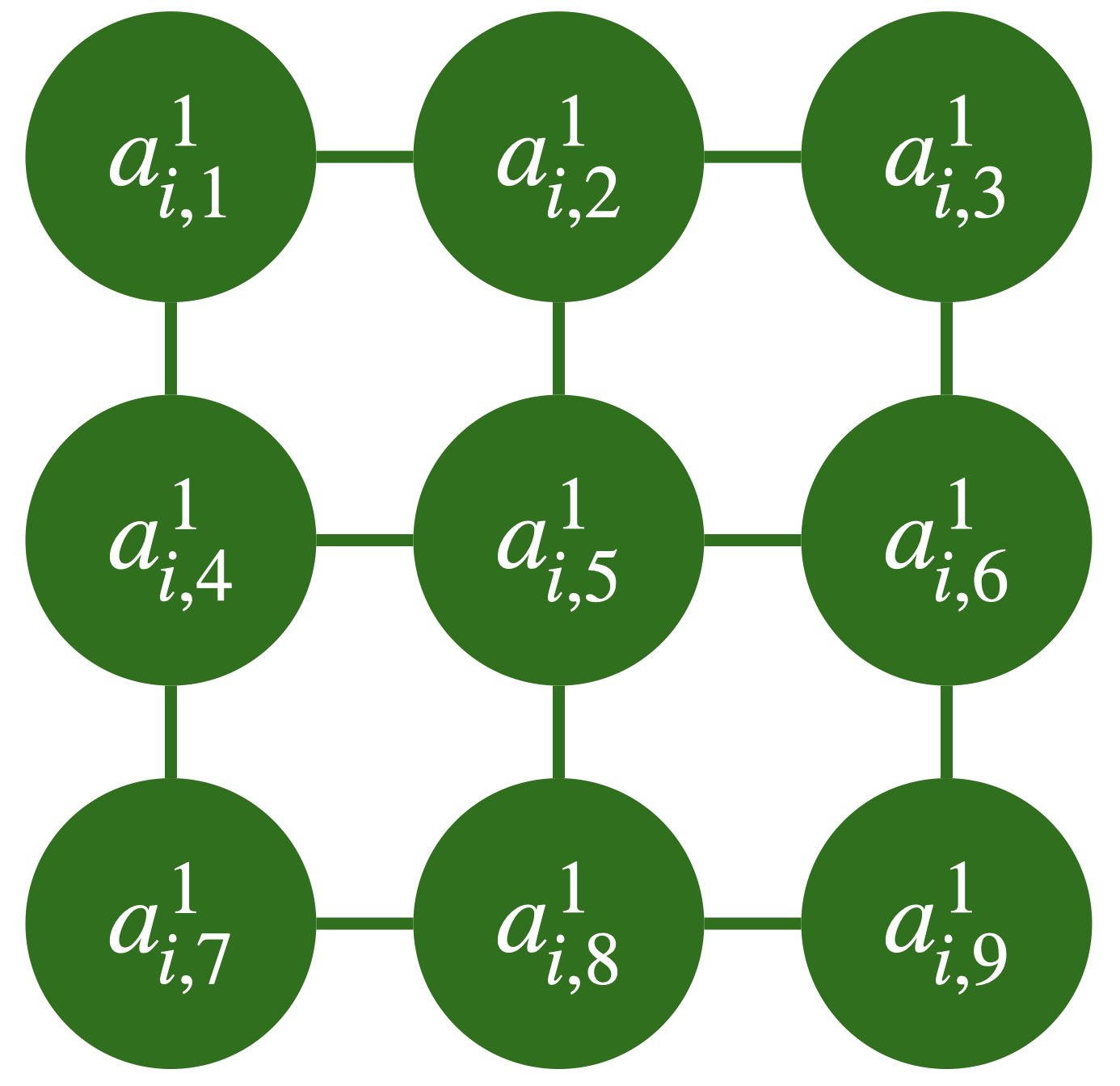}
        \caption{Atmospheric parameter graph for a $3\times3$ spatial grid.}
        \label{fig:graph_9nodeatm}
    \end{subfigure}
    \hfill
    \begin{subfigure}[b]{0.49\linewidth}
        \centering
        \includegraphics[width=0.5\linewidth]{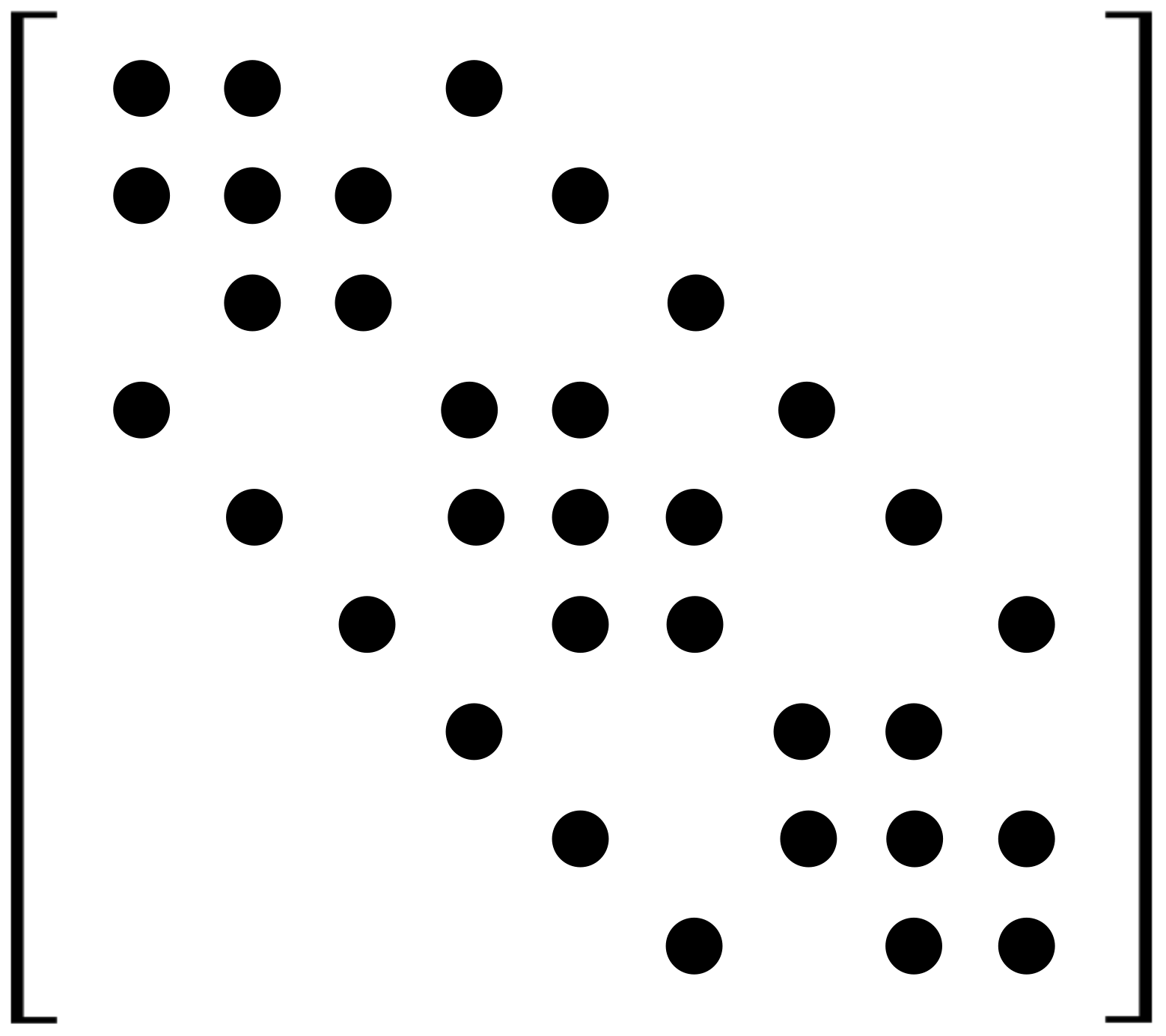}
        \caption{Sparsity structure of $\Sigma_{a_i}$. Each dot represents a non-zero scalar value.}
        \label{fig:sparsity_atm}
    \end{subfigure}
    \caption{Graphical structure and corresponding covariance sparsity for atmospheric parameters.}
    \label{fig:atm_graph_sparsity}
\end{figure}

Atmospheric parameters are connected to their four spatial neighbors but independent across parameters, so each $a_i$ is treated separately. For a $3\times 3$ grid $(m=9)$, Figure~\ref{fig:atm_graph_sparsity} shows the resulting pentadiagonal precision structure with edge effects. Each atmospheric parameter has a sparse precision $\Sigma_{a_i}^{-1} \in \mathbb{R}^{m \times m}$.



\begin{figure}
    \centering
    \includegraphics[width=0.3\linewidth]{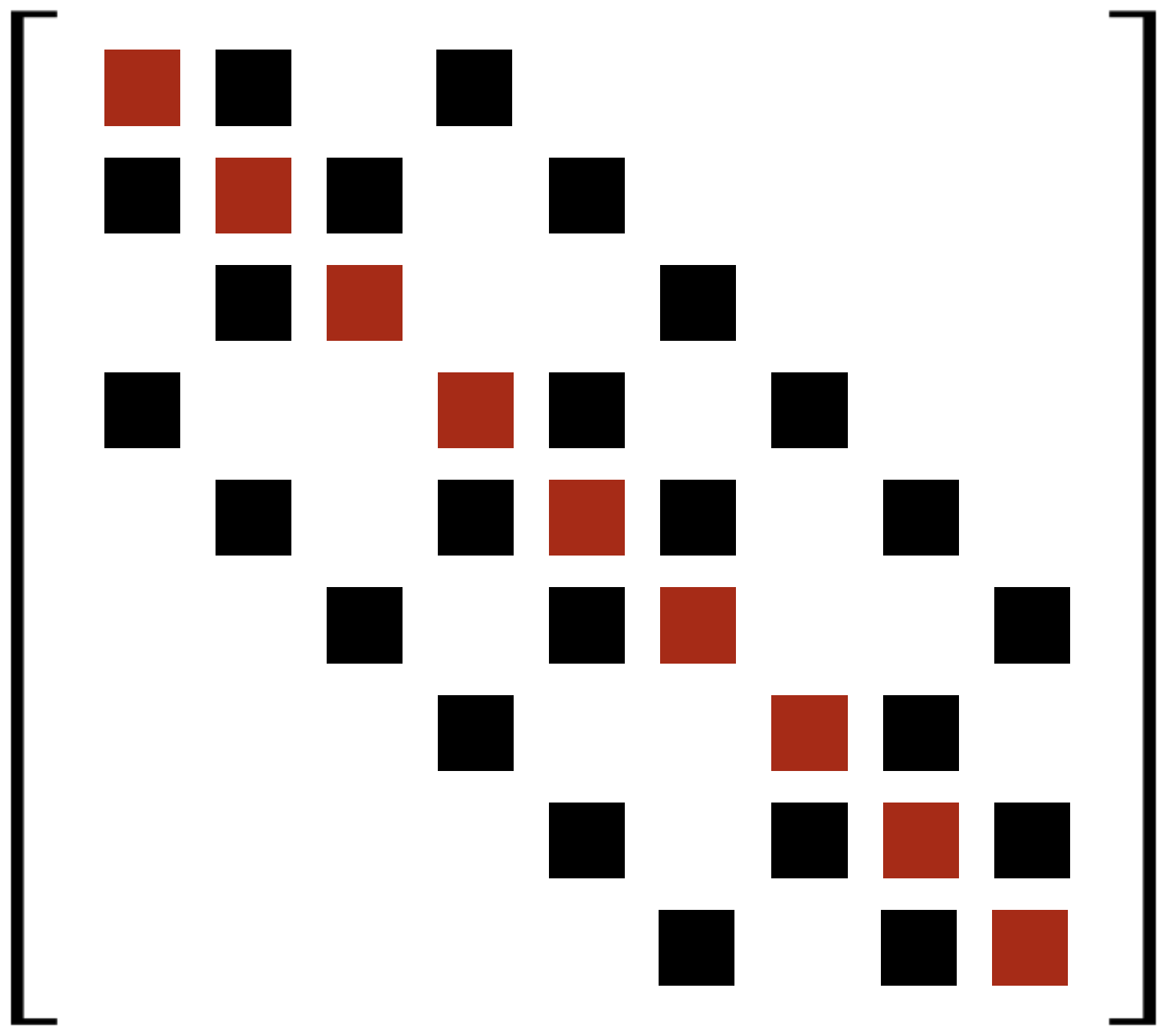}
    \caption{Sparsity structure of $\pmb{\Lambda}$. Red squares are non-zero diagonal $n\times n$ blocks; black squares are non-zero off-diagonal $n\times n$  blocks.}
    \label{fig:sparsity_prec}
\end{figure}

The joint prior precision $\pmb\Lambda \in \mathbbm{R}^{nm \times nm}$ is assembled from the surface and atmospheric precisions.
Since spatial correlations arise only from atmospheric parameters, the sparsity pattern of $\pmb\Lambda$ follows that of $\Sigma_{a_i}^{-1}$, as shown in Figure~\ref{fig:sparsity_prec}.

For spatial indices $j, k = 1,\dots,m$, the block structure of $\pmb\Lambda$ is:
\begin{align}
    \Lambda_{jj} &= \text{diag}\left( \Gamma_{s_j}^{-1}, \lambda_{1,jj}, \dots,  \lambda_{n_a,jj} \right)\\
    \Lambda_{jk} &=  \text{diag}\left(  \mathbf{0}, \lambda_{1,jk}, \dots,  \lambda_{n_a,jk} \right),\,\, j \neq k
\end{align}
where $\lambda_{i,jj} = [\Sigma_{a_i}^{-1}]_{jj}$ is the diagonal entry of the atmospheric precision for $a_i$ at location $j$, and $\lambda_{i,jk} = [\Sigma_{a_i}^{-1}]_{jk}$ is the corresponding off-diagonal element $(j,k)$.
Diagonal blocks have a dense surface component  ($\Gamma_{s_j}$ is generally dense) and a diagonal atmospheric component (the $n_a$ atmospheric parameters are mutually independent).

This structure enables significant computational savings: the full precision matrix is compactly stored as $m$ local surface covariances $\Gamma_{s_j}$ and $n_a$ sparse matrices $\Sigma_{a_i}^{-1}$, reducing storage from $\mathcal{O}(n^2 m^2)$ to $\mathcal{O}(n_s^2 m + n_a m)$. We show in the following subsections how this structure is preserved through both the inference and forecast steps of filtering.


\subsection{Bayesian Inference}

In the Bayesian inference step, observations at time $t$ update the state estimate.
Standard filtering algorithms invert the state covariance matrix at this step; our approach avoids this by operating directly on the precision matrix.


Since the forward model is nonlinear, we approximate the posterior mean via MAP estimation and the covariance via a Laplace approximation centered at the MAP, yielding a Gaussian that can be propagated forward in time. The MAP is found by minimizing the negative log posterior:
\begin{align}
    \mu^{t|t} \approx \mathbf{x}_{1:m}^* &=  \min_{\mathbf{x}_{1:m}} Q(\mathbf{x}_{1:m}^t; \mathbf{y}_{1:m}^t),
\end{align}
where the objective $Q$ decomposes as (derivation in Appendix~\ref{sec:simplifymapcost}):
\begin{align}
    Q(\mathbf{x}_{1:m}; \mathbf{y}_{1:m}) &= \sum_{j = 1}^m  \left( \| \mathbf{y}_{j} - f(\mathbf{x}_j) \|^2_{\Gamma_\epsilon} + \|\mathbf{s}_j - \mu_s \|^2_{\Gamma_{s_j}} \right) +  \sum_{i=1}^{n_a} \|\mathbf{a}_{i,1:m} - \mu_{a_i} \|^2_{\Sigma_{a_i}}.
\end{align}
This decomposition reflects three information sources: the likelihood, the spatially independent surface prior, and the spatially coupled atmospheric prior. Prior sparsity means the formulation involves only $n \times n$ and $m \times m$ blocks rather than a full $nm \times nm$ matrix.

We minimize $Q$ using L-BFGS, which uses gradients to construct low-rank approximations to the inverse Hessian. Our setup allows for analytical gradients:
\begin{align}
    \frac{\partial Q}{\partial \mathbf{s}_j} &=\left( \frac{\partial f}{\partial \mathbf{s}_j} \right)^\top \Gamma_\epsilon^{-1} (\mathbf{y}_j - f(\mathbf{s}_j, \mathbf{a}_j)) + \Gamma_s^{-1} (\mathbf{s}_j - \mu_s) \\
    \frac{\partial Q}{\partial a_{i,j}} &=  \left( \frac{\partial f}{\partial a_{i,j}} \right)^\top \Gamma_\epsilon^{-1} (\mathbf{y}_j - f(\mathbf{s}_j, \mathbf{a}_j)) \\
    &+ [\Lambda_{a_i}]_{jj}(a_{i,j} - \mu_{a_i}) + \frac{1}{2} \sum_{k \in N(j)} [\Lambda_{a_i}]_{jk} (a_{i,k}- \mu_{a_i}). \nonumber
\end{align}

We further use a block-coordinate optimization strategy alternating between surface and atmospheric parameters.
When atmospheric parameters are fixed, the surface-to-radiance mapping is nearly linear, so the surface subproblem over $n_s m$ parameters converges quickly. When surface parameters are fixed, optimization reduces to $n_a m$ atmospheric parameters. 
Because the surface-to-radiance mapping is nearly linear, the surface subproblem typically converges quickly even with a larger number of parameters. 
In contrast, the atmospheric component is more nonlinear and sensitive, making it harder to optimize. 
This alternating scheme exploits partial separability and yields more stable convergence, particularly when  $n_a \ll n_s$.

While sparsity benefits the MAP estimation, its full utility is realized in the Laplace approximation. 
In classic Kalman filtering, the posterior covariance requires explicit inversion. 
In contrast, in the information form, the Laplace approximation simplifies to an additive update to the precision:

\begin{align}
\label{eq:laplace}
    \pmb{\Lambda}^{t|t} = \pmb{\Lambda}^{t|t-1} +\nabla F(\mathbf{x}_{1:m}^{*\,t}) \,  \Lambda_\epsilon \, {\nabla F(\mathbf{x}_{1:m}^{*\,t})}^\top,
\end{align}
where $\mathbf{x}_{1:m}^{*\,t}$ is the MAP estimate at time $t$.

The additive term introduces new couplings between variables within each spatial block due to the forward model.
However, these couplings are local and no new connections are introduced between different spatial locations.
This is because the observation noise precision $\Lambda_\epsilon$ is block diagonal, reflecting the assumption that measurement noise is uncorrelated across pixels.
Moreover, the Jacobian $\nabla F$ is block-diagonal:
\begin{align}
    \nabla F(\mathbf{x}_{1:m}) = \left[\nabla f(\mathbf{s}_1, \mathbf{a}_1)^\top , \dots , \nabla f(\mathbf{s}_m, \mathbf{a}_m)^\top \right]^\top,
\end{align}
where each $\nabla f(\mathbf{s}_j, \mathbf{a}_j) \in \mathbbm{R}^{n_y \times n}$ is generally dense.
Although the within-block precision matrices $\Lambda_{jj}$ become dense due to the Jacobians, the global precision matrix $\pmb{\Lambda}^{t|t}$ retains its block-sparse structure.
This means that while inference at each pixel now involves dense coupling between the surface and the atmosphere, the overall filtering problem remains scalable due to the absence of new cross-pixel terms.
Thus, the structural sparsity of the precision continues to yield computational benefits in the inference step.

\subsection{Forecasting}

In the forecast stage at each time step, we propagate the state distribution forward using the dynamical model.
In our graphical model, only the surface parameters have temporal dependencies, and the atmospheric parameters are assumed independent across time.
Moreover, surface updates are performed independently at each spatial pixel.

To advance the surface distribution from $t-1$ to $t$, we apply the update equations in \eqref{eq:dynsystem}.
Assuming Gaussianity, the updates for the mean and covariance at location $j$ are:
\begin{align}
    \mu_{s_j}^{t+1|t} &= \alpha \mu_{s_j}^{t|t} + (1-\alpha) \mu_{s_j}^1 \nonumber \\
    \Gamma_{s_j}^{t+1|t} &= \alpha^2 \Gamma_{s_j}^{t|t} + (1-\alpha)^2 \beta^2 \Gamma_{s_j}^1,
    \label{eq:covforecast_s}
\end{align}
where $\mu_{s_j}^1$ and $\Gamma_{s_j}^1$ are the initial mean and covariance of the surface at location $j$.
The hyperparameter $\alpha$ controls the weighting between the posterior and prior means to prevent overconfidence, while $\beta$ inflates the covariance for added flexibility.

After computing $\Gamma_{s_j}^{t+1|t}$, we update the corresponding diagonal blocks in the prior precision $\pmb{\Lambda}^{t+1|t}$ with $(\Gamma_{s_j}^{t+1|t})^{-1}$.
Since the graphical model structure is preserved, $\pmb{\Lambda}^{t+1|t}$ remains sparse.
Only the surface-related blocks are modified, whereas the atmospheric precision remains unchanged across time.

\subsubsection{Marginals of the Joint Posterior}

The challenge in applying the forecast update lies in the transition from joint to marginal representations.
The inference step produces the joint mean $\mu^{t|t}$ and precision $\pmb{\Lambda}^{t|t}$, but the forecast update is defined over the marginal surface distributions at each location.

The marginal mean and covariance at pixel $j$ are extracted from the joint posterior as:
\begin{align}
    \mu_{s_j}^{t|t} = \left[ \left[\mu^{t|t} \right]_j\right]_s, \qquad
    \Gamma_{s_j}^{t|t} = \left[\left[ \left(\pmb\Lambda^{t|t}\right)^{-1} \right]_{jj}\right]_{ss}.
\end{align}


Directly computing $\pmb\Lambda^{-1}$ would eliminate sparsity and dramatically increase storage and computational costs.
To avoid this, we approximate the marginal using a localized inference technique inspired by belief propagation.

We are interested in extracting the surface marginal covariances from the full posterior precision matrix $\pmb{\Lambda}^{t|t}$, specifically the surface sub-blocks of the diagonal elements $[\pmb{\Lambda}^{-1}]_{jj}$ for $j = 1, \dots, m$.
This task is nontrivial because the spatial structure of the graphical model induces dependencies across pixels, meaning the marginal at a single location depends on the state at neighboring locations, which complicates direct extraction from the joint precision matrix.

To illustrate the concept behind our approximation, consider a joint Gaussian distribution over two disjoint sets of variables $\mathbf{x}_A$ and $\mathbf{x}_B$, where
$p(\mathbf{x}_A, \mathbf{x}_B) = \mathcal{N}(\mathbf{0}, M^{-1})$, and the precision matrix $M$ is partitioned as:
\begin{align}
    M = \begin{bmatrix}
        M_{AA} & M_{AB}\\
        M_{AB}^\top & M_{BB}
    \end{bmatrix}.
\end{align}
Then, the marginal distribution over $\mathbf{x}_A$ is also Gaussian: 
\begin{align}
    p(\mathbf{x}_A) = \int p(\mathbf{x}_A, \mathbf{x}_B) \, d\mathbf{x}_B = \mathcal{N}(\mathbf{0},\Sigma_A),
\end{align}
where the marginal covariance is given by the inverse of the Schur complement:
\begin{align}
    \Sigma_A = (M_{AA} - M_{AB} M_{BB}^{-1} M_{AB}^\top)^{-1}.
\end{align}

We apply this principle to extract the marginal surface covariance for location $j$, denoting $A = \{j\}$ and $B = \{1, \dots, m\} \setminus \{j\}$.
The Schur complement expression becomes:
\begin{align}
    \left[ \pmb{\Lambda}^{-1} \right]_{jj} = \left( \pmb{\Lambda}_{jj} - \pmb{\Lambda}_{jB} \left(\pmb{\Lambda}_{BB}\right)^{-1} \pmb{\Lambda}_{jB}^\top \right)^{-1}.
\end{align}
This formulation allows us to bypass computing the full inverse $\pmb{\Lambda}^{-1}$, but it still requires inverting $\pmb{\Lambda}_{BB} \in \mathbb{R}^{n(m-1) \times n(m-1)}$, which is expensive when $m$ is large.

\subsubsection{Approximating the Marginals}
\label{sec:approximatemarginals}

We use the belief propagation algorithm as conceptual motivation for our approximation.
To compute the marginal of $\mathbf{x}_j$, belief propagation integrates out variables in $B$ in a sequence that respects the conditional dependencies in the graph, often proceeding from the most distant nodes toward the immediate neighbors of $j$.
Each integration step is commonly interpreted as a \textit{message} passed along graph edges, where the message represents the update to a node’s belief given new information, which is the updated state from the inference stage at time $t$ in our case.
Since the nodes are connected through spatial dependencies in the atmospheric parameters, this message-passing process reflects the way state updates propagate through the graph toward $\mathbf{x}_j^t$.

\begin{figure}
\centering
\begin{subfigure}{.5\textwidth}
  \centering
  \includegraphics[width=0.4\linewidth]{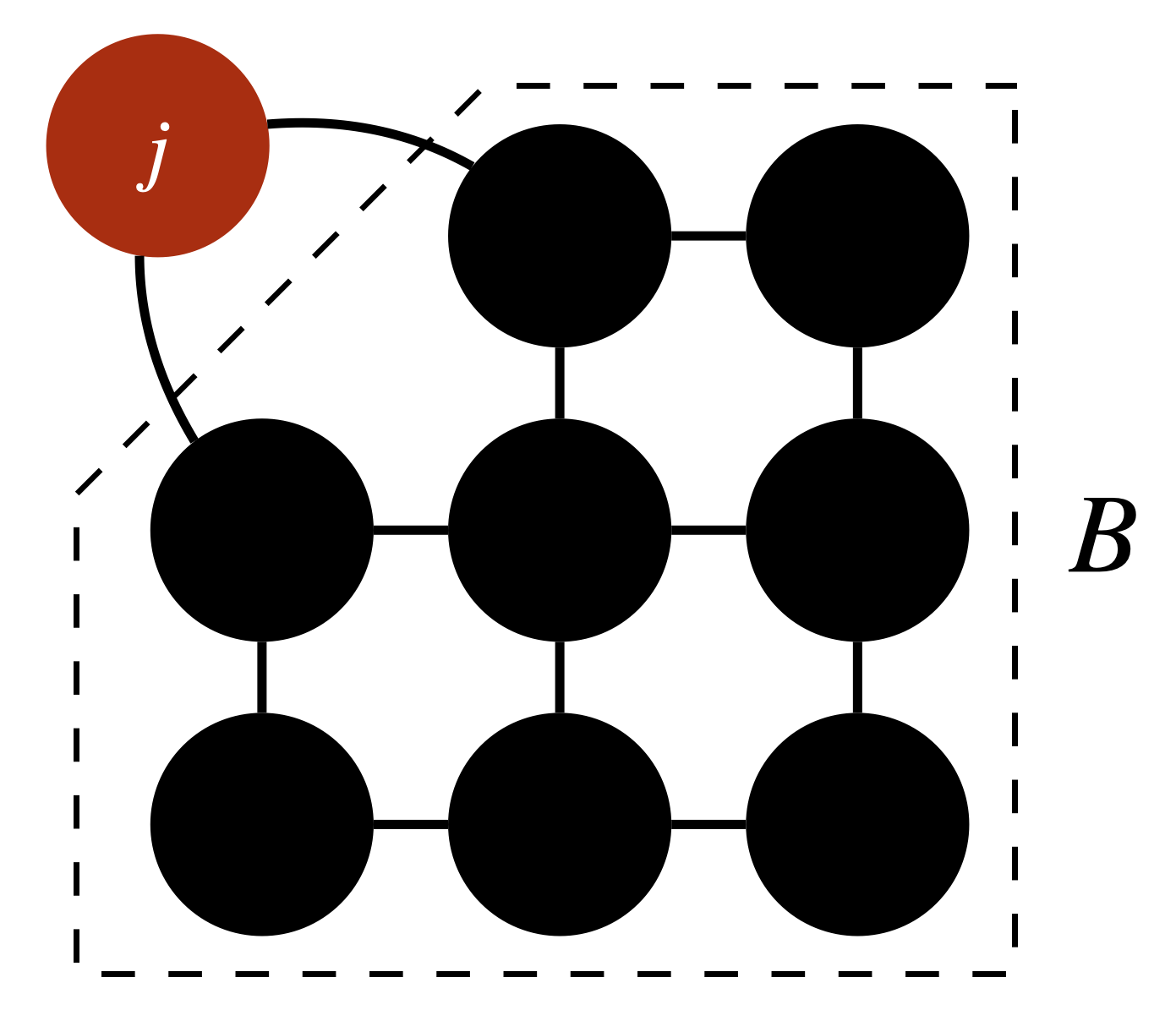}
  \caption{Grid structure of spatial field}
  \label{fig:marginal_approx_graph_left}
\end{subfigure}%
\begin{subfigure}{.5\textwidth}
  \centering
  \includegraphics[width=0.4\linewidth]{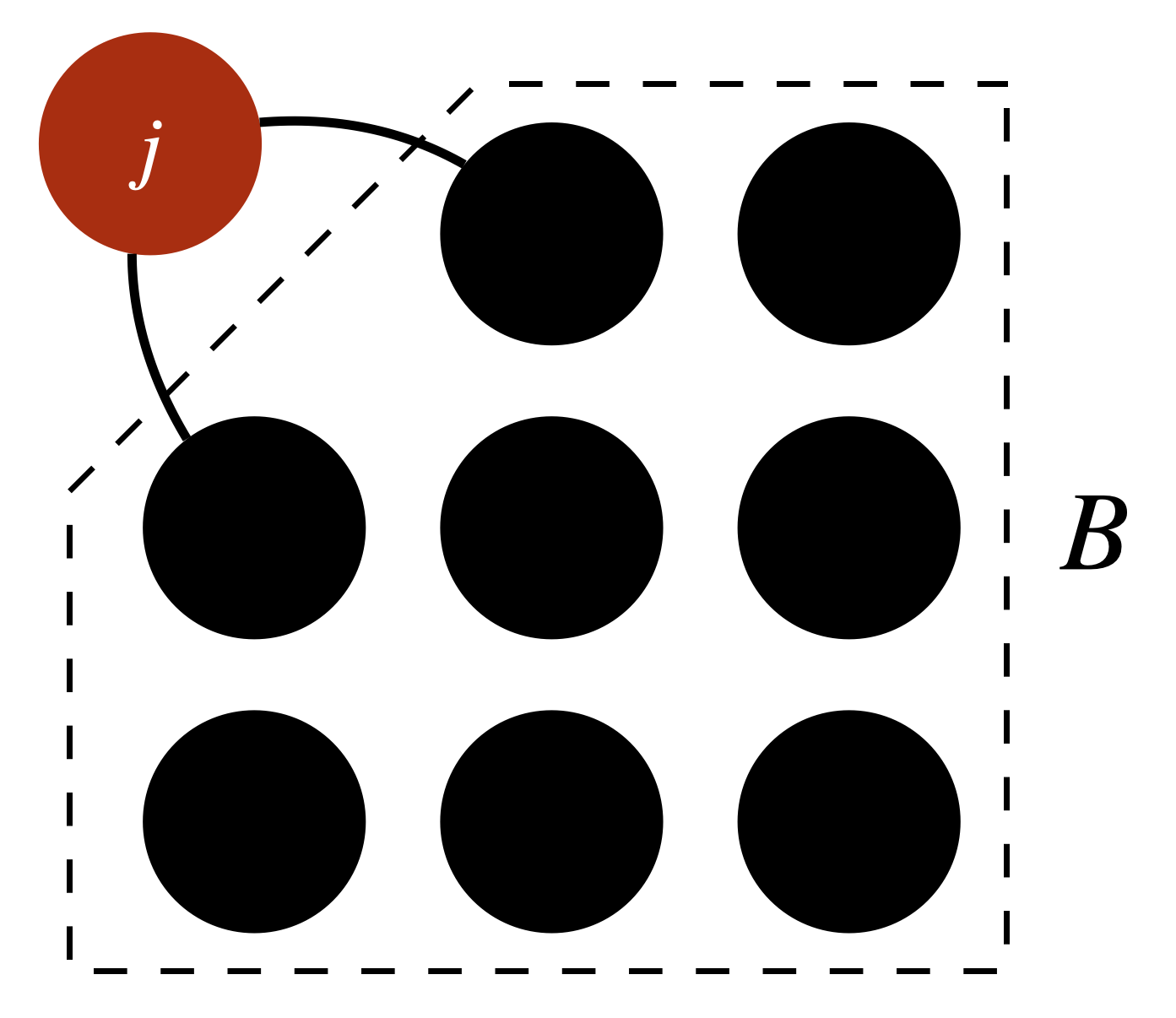}
  \caption{Independence assumption of nodes in $B$}
  \label{fig:marginal_approx_graph_right}
\end{subfigure}
\caption{Graphical model visualization of the approximation of the marginal of node $j$.}
\label{fig:marginal_approx_graph}
\end{figure}

In the general case, the grid graph in Figure~\ref{fig:marginal_approx_graph_left} corresponds to a precision matrix $\pmb\Lambda_{BB}$ with multiple off-diagonal blocks.
Inverting this matrix leads to a dense representation with computational complexity $\mathcal{O}(n^2m^2)$.

To reduce this cost, we assume that the most significant contributions to the marginal of $\mathbf{x}_j$ arise from its immediate neighbors.
We therefore simplify the graph by removing edges to non-neighboring nodes, leading to the simplified structure shown in Figure~\ref{fig:marginal_approx_graph_right}.
In this approximation, the nodes in $B \setminus {N(j)}$ are treated as mutually independent, which renders $\pmb{\Lambda}_{BB}$ block diagonal and easily invertible.

The marginal covariance of block $(j,j)$ then simplifies to:
\begin{align}
     \left[ \pmb{\Lambda}^{-1} \right]_{jj} \approx \left( \pmb{\Lambda}_{jj} - \sum_{k \in N(j)} \pmb{\Lambda}_{jk} \left(\pmb{\Lambda}_{kk}\right) ^{-1} \pmb{\Lambda}_{jk}^\top \right)^{-1}.
\end{align}

This reduces computational complexity from $\mathcal{O}(n^2 m^2)$ to $\mathcal{O}(n^2 k)$, where $k$ is the number of neighbors of node $j$ (typically at most four in a grid).
Empirically, we verified that this approximation yields a marginal covariance with an $L^2$-norm error of 1–2\% compared to the true marginal in a synthetic problem with $n_s = 3$ and $m = 100$ (see Appendix~\ref{sec:matrixinverror}).


Finally, applying this approximation to the remote sensing setting, the surface covariance for location $j$ is extracted from the marginal as:
\begin{align}
    \label{eq:surf_marginal}
    \Gamma_{s_j}^{t|t} \approx \left[ \left( \pmb\Lambda^{t|t}_{jj} - \sum_{k\in N(j)}\pmb\Lambda^{t|t}_{jk}  \left( \pmb\Lambda^{t|t}_{kk} \right)^{-1} \left( \pmb\Lambda^{t|t}_{jk} \right)^\top  \right)^{-1} \right]_{ss}.
\end{align}

\subsection{Practical Considerations and Initialization}

The methodology outlined in this section uses an arbitrary initialization of the precision matrix that adheres to sparsity structure implied by the conditional independence assumptions. 
In our numerical experiments, we initialize the problem using the same procedure as the nearest-neighbor Gaussian process (NNGP) \cite{datta_nearest-neighbor_2016}.
This initialization is applied independently to each atmospheric parameter.

To enforce the spatial smoothness in the atmospheric parameters, we prescribe a Gaussian process prior for each parameters using a squared exponential kernel with length scale $l$ and variance $\sigma_a^2$:
\begin{align}
    \label{eq:sqexpkernel}
    K^\text{sqexp}_i(z,z') = \sigma_a^2 \exp\left(-\frac{\|z - z'\|}{2l^2} \right),
\end{align}
where $(z,z')$ is the spatial location.

We begin with the Gaussian distribution $p_0 (\mathbf{a}_{i,1:m}) = N\left(\mu_{a_i}, \mathbf{K}^\text{sqexp}_i\right)$ for $i = 1\dots n_a$,
where $\left[\mathbf{K}^\text{sqexp}_i\right]_{jj'} = K^\text{sqexp}_i(j,j')$.
We then use the Vecchia approximation \cite{vecchia_estimation_1988} to enforce the grid graphical structure that is desired in the atmosphere.
\begin{align}
    p_0 (a_{i,1:m}) = p_0 (a_{i,1}) \, p_0 (a_{i,2} | a_{i,1}) \dots  p_0 (a_{i,2} | a_{i,1}) \approx \prod_{j=1}^m p_0 (a_{i,j} | a_{i,N(j)}) ,
\end{align}
where $N(j)$ are the neighbors of $j$ according to the graphical model. 
Since a grid structure is assigned to the graph, each node has a maximum of four neighbors. 

Each conditional $a_{i,j} | a_{i,N(j)}$ is distributed according to:
\begin{align}
    p_0 (a_{i,j} | a_{i,N(j)}) = N\left(\mathbf{A}_j a_{i,N(j)}, \mathbf{B}_j\right),
\end{align}
where $\mathbf{A}_j = \mathbf{K}_{j,N(j)} \mathbf{K}_{N(j),N(j)}^{-1}$ and $\mathbf{B}_j = \mathbf{K}_{j,j} - \mathbf{K}_{j,N(j)} \mathbf{K}_{N(j),N(j)}^{-1} \mathbf{K}_{N(j),j}$.

The resulting prior for each atmospheric parameter $a_{i,1:m}$ is defined by the product of the conditional distributions:
\begin{align}
    p (a_{i,1:m})  = \prod_{j=1}^m p_0 (a_{i,j} | \mathbf{a}_{i,N(j)}) = N(\mu_{a,i}, \Sigma_{a,i}),
\end{align}
where $\mu_{a,i}$ and $\Sigma_{a,i}$ are the mean and covariance computed by combining the Gaussian conditionals in closed form.
This construction induces a sparse precision matrix $\Sigma_{a,i}^{-1}$, whose sparsity pattern matches the neighborhood structure defined by the graphical model on the spatial grid.

\subsection{Summary of Sparsity Patterns}

\begin{figure}
    \centering
    \includegraphics[width=\linewidth]{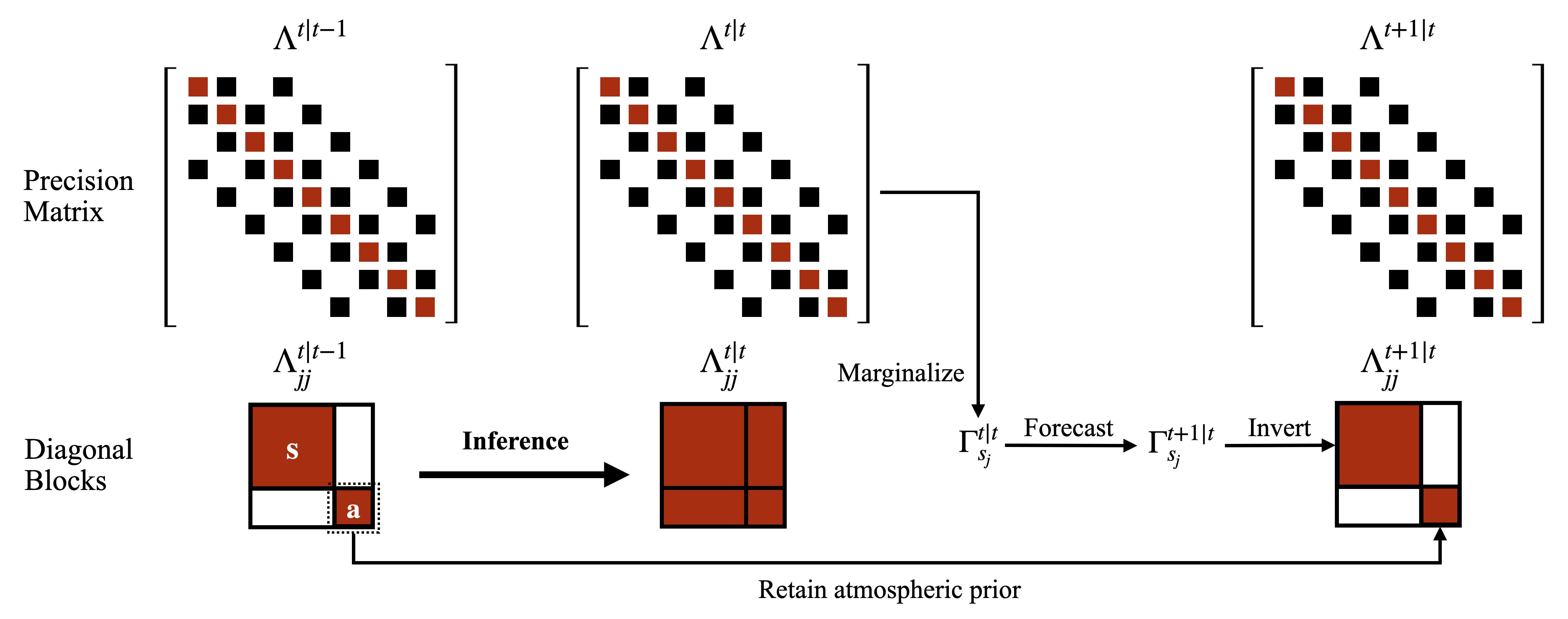}
    \caption{Progression of sparsity throughout one time step of the algorithm. The spatial structure remains the same throughout.}
    \label{fig:sparsity_progression}
\end{figure}

Figure~\ref{fig:sparsity_progression} illustrates the progression of sparsity in the precision matrix through one full filter time step. The top row shows the block sparsity structure over the spatial domain, which is dictated by conditional independence and remains fixed throughout. The bottom row shows how each diagonal block evolves during inference and forecasting.

Starting from the prior at time $t$, the inference step introduces dense coupling between surface and atmospheric parameters at each location via the Laplace update~\eqref{eq:laplace}, densifying the diagonal blocks while preserving the overall block sparsity. To advance in time, approximate marginal covariances of the surface parameters $\mathbf{s}_j$ are extracted using the Schur complement approximation~\eqref{eq:surf_marginal}, then propagated forward via~\eqref{eq:covforecast_s}. The resulting forecasted surface covariances are inverted and inserted into the surface subblocks of the diagonal precision blocks. Atmospheric subblocks remain unchanged, since atmospheric parameters are temporally independent.


This structure allows the algorithm to scale efficiently by avoiding unnecessary updates to time-invariant portions of the precision matrix while maintaining a consistent sparsity pattern throughout.

\section{Numerical Experiments}
\label{sec:results}

We evaluate the proposed filtering algorithm on two remote sensing applications to demonstrate its generality and effectiveness.

The first is based on NASA's OCO-2 mission \cite{eldering_orbiting_nodate}, where the state vector consists of surface and atmospheric parameters, with the atmospheric component dominated by CO$_2$ concentrations at various altitudes. 
The quantity of interest is the column-averaged CO$_2$, a weighted average over the vertical profile. 
We use a linearized forward model with synthetically generated spatio-temporal data, and show that filtering over the surface state can nonetheless improve CO$_2$ retrieval accuracy.

The second example targets surface reflectance retrievals from visible to shortwave infrared (VSWIR) imaging spectroscopy, aligned with NASA's EMIT and EAGLE missions. 
The quantity of interest is the surface reflectance spectrum, with atmospheric parameters also retrieved. 
The forward model is nonlinear, and we use real time series data collected from an airborne imaging spectrometer, AVIRIS-NG. 
This example highlights the algorithm’s ability to improve surface retrievals using real data and a nonlinear forward model, focusing specifically on estimating surface parameters over time.

Across both experiments, we compare three retrieval strategies. 
The \textit{independent} retrieval models no spatial or temporal interactions; each pixel at each time step is retrieved independently using a common prior with surface covariance $\Gamma_s$ and atmospheric variances $\sigma_{a_i}^2$.
The \textit{spatial} retrieval incorporates atmospheric spatial correlations via prior $(\mu^1, \pmb\Lambda^1)$ but without cross-time information exchange. The \textit{spatio-temporal} retrieval models both structures, sequentially updating the spatial prior via the forecast and inference steps of Section~\ref{sec:methodology}.


\subsection{Orbiting Carbon Observatory-2}

The Orbiting Carbon Observatory-2 (OCO-2) mission was launched in 2014 and designed to provide precise, global measurements of atmospheric carbon dioxide (CO$_2$).
The radiance is measured in three narrow bands: the oxygen A-band ($\sim$0.77 µm), a weak CO$_2$ absorption band ($\sim$1.61 µm), and a strong CO$_2$ band ($\sim$2.06 µm).
These repeated-orbit observations naturally form a spatio-temporal data structure, with total column CO$_2$ ($XCO_2$) as the primary quantity of interest. We build on the Lamont 2015 case study of \cite{hobbs_spatial_2021}, adopting their experimental setup as a common reference point and extending it with our spatio-temporal filtering algorithm.


We use a reduced-order state vector of 36 variables (Table~\ref{tab:oco2_state}): $n_s = 3$ surface reflectance parameters across the three spectral bands, and $n_a = 33$ atmospheric variables including a 20-level CO$_2$ vertical profile, surface pressure, and a set of aerosol properties for four types (dust, ice, water, and sulfate).
$XCO_2$ is a linear projection of the atmospheric state:
\begin{align}
    XCO_2 = \mathbf{w}^\top \mathbf{a},
\end{align}
where $w_i > 0$ only for $i = 1, \dots, 20$. 
\begin{table}
    \centering
    \begin{tabular}{|c|c|c|}
        \hline
        Collection & Length & Description\\
        \hline
        Surface Albedo & 3 & Parameters describing surface reflectance\\
        CO$_2$ Vertical Profile & 20 & CO$_2$ concentration at 20 specified altitudes\\
        Surface Pressure & 1 & Used as a reference point for altitudes\\
        Aerosols & 12 & Parameters pertaining to four aerosol types\\
        \hline
    \end{tabular}
    \caption{OCO-2 state vector components.}
    \label{tab:oco2_state}
\end{table}

The forward model is linearized around the atmospheric state retrieved near Lamont, Oklahoma (~97°W, 37°N) during the October 2015 overpass, giving $f(\mathbf{x}) = G \mathbf{x}$ and closed-form posterior expressions:
\begin{align}
    \pmb{\Lambda}^{t|t} &= \pmb{\Lambda}^{t|t-1} + G^\top \Lambda_\epsilon G\\
    \mu^{t|t} &= \left( \pmb{\Lambda}^{t|t}\right)^{-1} \left( G^\top \Lambda_\epsilon \mathbf{y}^t + \pmb{\Lambda}^{t|t-1} \mu^{t|t-1} \right).
\end{align}

\subsubsection{Simulated Dataset}

We simulate a time series over a $20\times 20$ spatial grid ($m=400$ locations) for $T = 10$ time steps.

Atmospheric states are drawn from a Gaussian process with squared exponential kernel \eqref{eq:sqexpkernel}, with full covariance constructed as the Kronecker product of the within-state covariance $\Gamma_a$ and spatial kernel matrix $\mathbf{K}$:
\begin{align}
    \mathbf{a}_\text{true}^t \sim \mathcal{N} (\text{vec}(\mu_a  \mathbf{1}_m^\top), \Gamma_a \otimes \mathbf{K}),
\end{align}
with fixed length scale $l = 10$ and set $\gamma_a = 1$.
Surface reflectance evolves via an autoregressive model:
\begin{align}
    \mu_s^t = \alpha_s \mu_s^{t-1} + (1-\alpha_s) \beta_s \mu_s - \gamma_s + \epsilon_s
\end{align}
with $\alpha_s = 0.5$, $\beta_s = 1.3$, $\gamma_s = 0.005$, and $\epsilon_s \sim U(-0.01, 0.01)$. 
True surface states are drawn from $N(\mu_s^t, \gamma_s \Gamma_s)$ with $\gamma_s = 0.04$. 
Observations are generated by applying the forward model to the true state:
\begin{align}
    \mathbf{y}_\text{obs} = F \mathbf{x}_\text{true} + \epsilon, \quad \epsilon \sim \mathcal{N}(0, \Gamma_\epsilon)
\end{align}
where $\Gamma_\epsilon$ is determined by instrument noise. Although the true atmospheric parameters exhibit cross-parameter correlations that our algorithm does not model, improvements in retrieval accuracy are still observed.

\subsubsection{Results}

Retrievals of the state were performed using the proposed filtering algorithm on simulated observations ${\mathbf{y}_\text{obs}}_{1:m}^{1:T}$ and compared against the known ground truth used to generate the data. 
The autoregressive weighting and covariance inflation hyperparameters were tuned to $\alpha = 0.8$ and $\beta = 5$, and a length scale of $l=5$ was used in the kernel used to generate the initial prior precision.

\begin{figure}
\centering
\begin{subfigure}{.5\textwidth}
  \centering
  \includegraphics[width=0.98\linewidth]{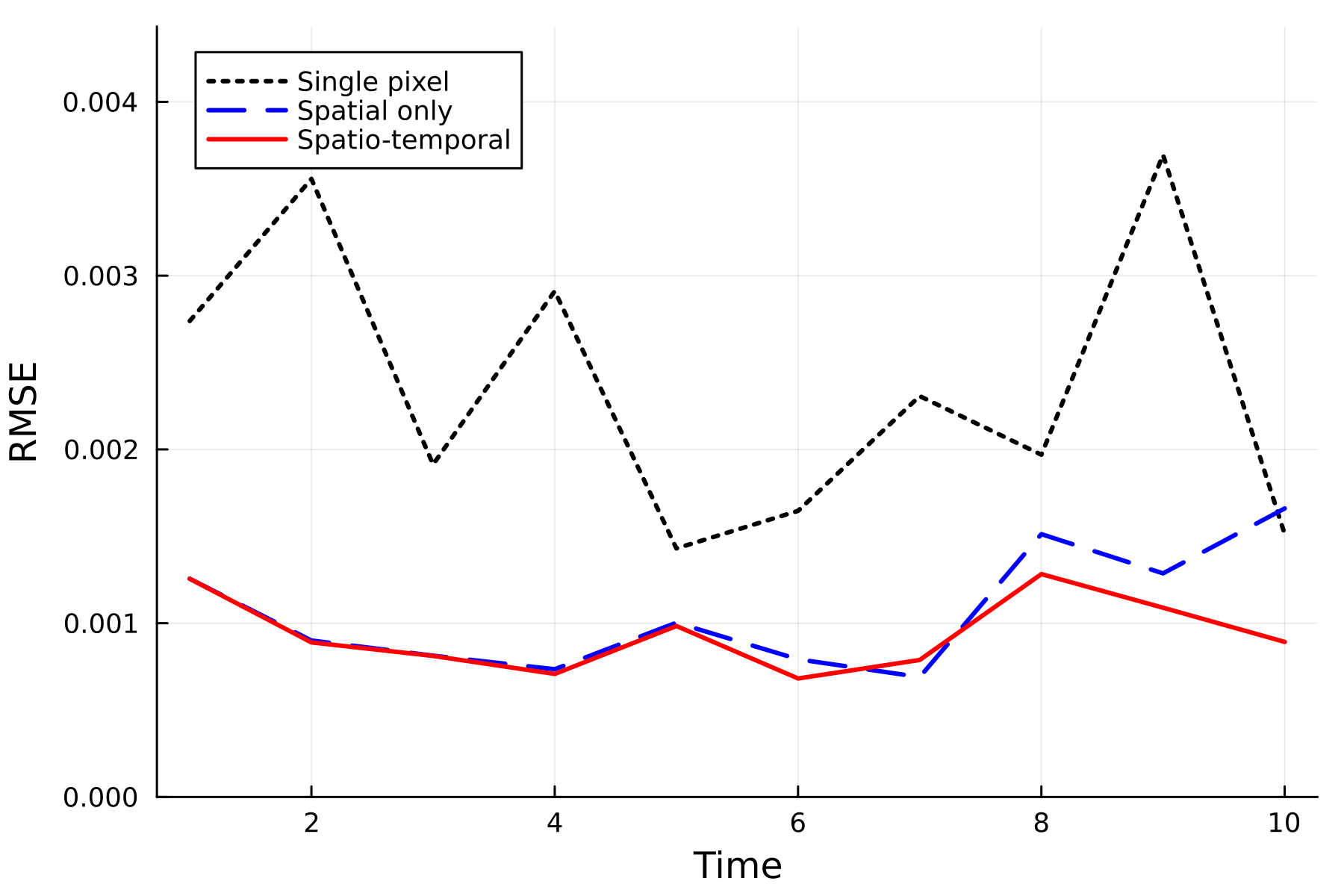}
  \caption{Error over time for surface parameters.}
  \label{fig:errorOverTime_oco2_surface}
\end{subfigure}%
\begin{subfigure}{.5\textwidth}
  \centering
  \includegraphics[width=0.98\linewidth]{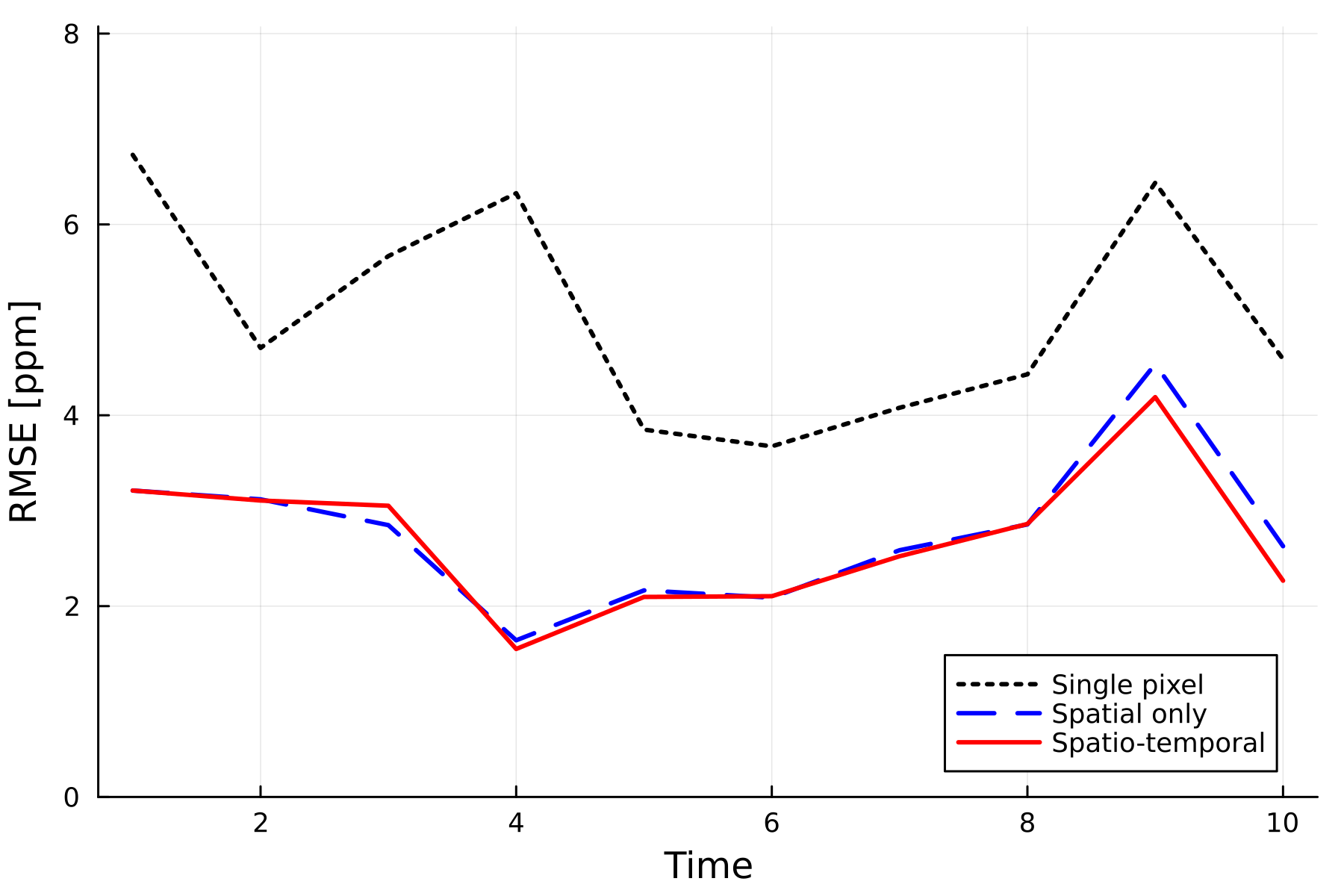}
  \caption{Error over time for $XCO_2$.}
  \label{fig:errorOverTime_oco2_xco2}
\end{subfigure}
\caption{OCO-2 Example. Root mean squared error over time. Spatio-temporal structure improves accuracy across both surface and atmospheric components.}
\label{fig:errorOverTime_oco2}
\end{figure}

Figure~\ref{fig:errorOverTime_oco2} presents the root mean squared error (RMSE) over time for both surface reflectance (left) and $XCO_2$ (right), averaged across all $m = 400$ spatial locations. 
In the independent (single-pixel) retrieval, errors fluctuate with time due to randomness in the measurements. 
Incorporating spatial structure (spatial retrieval) yields consistent reductions in error, and introducing temporal filtering (spatio-temporal retrieval) further improves the surface estimates.

While the primary quantity of interest is column-averaged CO$_2$ ($XCO_2$), we show both the error averaged over the three surface bands (Figure~\ref{fig:errorOverTime_oco2_surface}) and the error in $XCO_2$ (Figure~\ref{fig:errorOverTime_oco2_xco2}) to evaluate the impact on different components of the state. 
The spatial retrieval uses correlated atmospheric priors, which significantly reduces $XCO_2$ error compared to the baseline. 
This spatial structure also indirectly improves surface estimation. 
The spatio-temporal retrieval provides additional gains for surface parameters by propagating prior information through time using the filter. 
While this has a smaller effect on $XCO_2$, the improved surface estimates still yield a modest benefit.

Since there are relatively few surface parameters compared to atmospheric parameters in the OCO-2 state vector, the spatial correlations have a stronger effect on $XCO_2$ than the temporal correlations have on the surface.
This explains why the spatial retrieval performs nearly as well as the spatio-temporal retrieval in $XCO_2$ estimation. 

\subsection{Visible Shortwave Infrared Imaging Spectroscopy}

VSWIR imaging spectroscopy enables surface reflectance retrieval across the 380–2500 nm spectral range. NASA's EMIT mission focuses on monitoring surface mineral composition in arid regions \cite{thompson_-orbit_2024}, while the upcoming EAGLE mission is designed to quantify vegetation health, snow and ice cover, and soil chemistry \cite{cawse-nicholson_nasas_2021, cawse-nicholson_surface_2023}. These are part of a growing fleet of spaceborne imaging spectrometers, including NASA's PACE OCI, and international missions such as EnMAP, DESIS, PRISMA, and the upcoming ESA CHIME \cite{meister_ocean_2024, guanter_enmap_2015, krutz_instrument_2019, cogliati_prisma_2021, nieke_towards_2018}, that are rapidly expanding the availability of hyperspectral Earth observation data. These satellite instruments are supported by airborne spectrometers such as AVIRIS-NG \cite{green_2023_aviris}.


Unlike the OCO-2 experiment where atmospheric parameters dominate the state vector, VSWIR retrievals are surface-dominant, as most quantities of interest are directly derived from surface spectral signatures. We consider a state vector with $n_s = 326$ surface parameters and $n_a = 2$ atmospheric parameters, with $n_y = n_s = 326$ radiance bands.


The atmospheric priors, $p(a_i) = N(\mu_{a_i}, \sigma_{a_i})$ for $i \in \{1,\dots, n_a\}$, are chosen to be deliberately uninformative to encourage learning from the data. 
The surface prior is constructed from a library of historical reflectance spectra, clustered into eight representative surface types. Each cluster is modeled by a Gaussian distribution. 
For each pixel, the surface prior used in retrieval, $p (\mathbf{s}) = N(\mu_s, \Gamma_s)$, is selected by minimizing the Mahalanobis distance between the initial reflectance estimate and the cluster means.
The forward model describes light propagation from the surface to the spectrometer, with the $i$-th spectral band given by:
\begin{equation} \label{eq:fwd}
    f_i(s_i, \mathbf{a}) =  r_{i} (\mathbf{a}) + \frac{t_i(\mathbf{a}) \cdot s_i}{1 - \gamma_i (\mathbf{a}) \cdot s_i},
\end{equation}
where $r_i$, $t_i$, and $\gamma_i$ denote the path reflectance, transmittance, and spherical albedo – nonlinear black-box functions of the atmospheric state.


\subsubsection{Datasets}

We base our results in this section on operational radiance data obtained from the SBG High Frequency Time Series (SHIFT) campaign, a study used to emulate the EAGLE mission's time series data \cite{chadwick_unlocking_2025, brodrick_shift_2024}.
Radiance data is collected approximately weekly using AVIRIS-NG \cite{hamlin_imaging_2011}.
 Since per-pixel \textit{in situ} ground truth reflectance is impractical to collect, we designate the surface reflectance retrieved from AVIRIS-NG radiances as the ``true" surface reflectance. Synthetic atmospheric fields are generated at each time step and radiances are forward-simulated with additive noise.

Atmospheric fields are drawn from a hierarchical Gaussian process with no temporal correlation to emulate a realistic atmosphere. At each time step, a mean is sampled from a uniform distribution and a spatial field is drawn using a squared exponential kernel with correlation length $l = 10$.
Aerosol optical depth (AOD) means are sampled from $\text{Uniform}(0.1, 0.4)$, with covariance scaled by 0.01.
Water vapor (H$_2$O) means are sampled from $\text{Uniform}(1.2, 1.8)$, with covariance scaled by 0.05.
Table~\ref{tab:sim_data} summarizes the mean atmospheric conditions for each time step alongside the corresponding SHIFT flight dates.




\begin{table}[ht]
    \centering
    \footnotesize
    \begin{tabular}{|c|c|c|c|c|c|c|c|c|c|c|}
        \hline
        Timestep & 1 & 2 & 3 & 4 & 5 & 6 & 7 & 8 & 9 & 10 \\
        \hline
        Flight Date & 2/28 & 3/08 & 3/22 & 4/05 & 4/12 & 4/20 & 4/29 & 5/11 & 5/17 & 5/29 \\
        \hline
        AOD mean & 0.171 & 0.297 & 0.137 & 0.293 & 0.392 & 0.311 & 0.369 & 0.385 & 0.275 & 0.320 \\
        \hline
        H$_2$O mean & 1.403 & 1.527 & 1.318 & 1.593 & 1.717 & 1.393 & 1.497 & 1.495 & 1.380 & 1.632 \\
        \hline
    \end{tabular}
    \caption{VSWIR Example. Simulated aerosol optical depth (AOD) and water vapor (H$_2$O) values by timestep.}
    \label{tab:sim_data}
\end{table}

\begin{figure}
    \centering
    \includegraphics[width=0.7\linewidth]{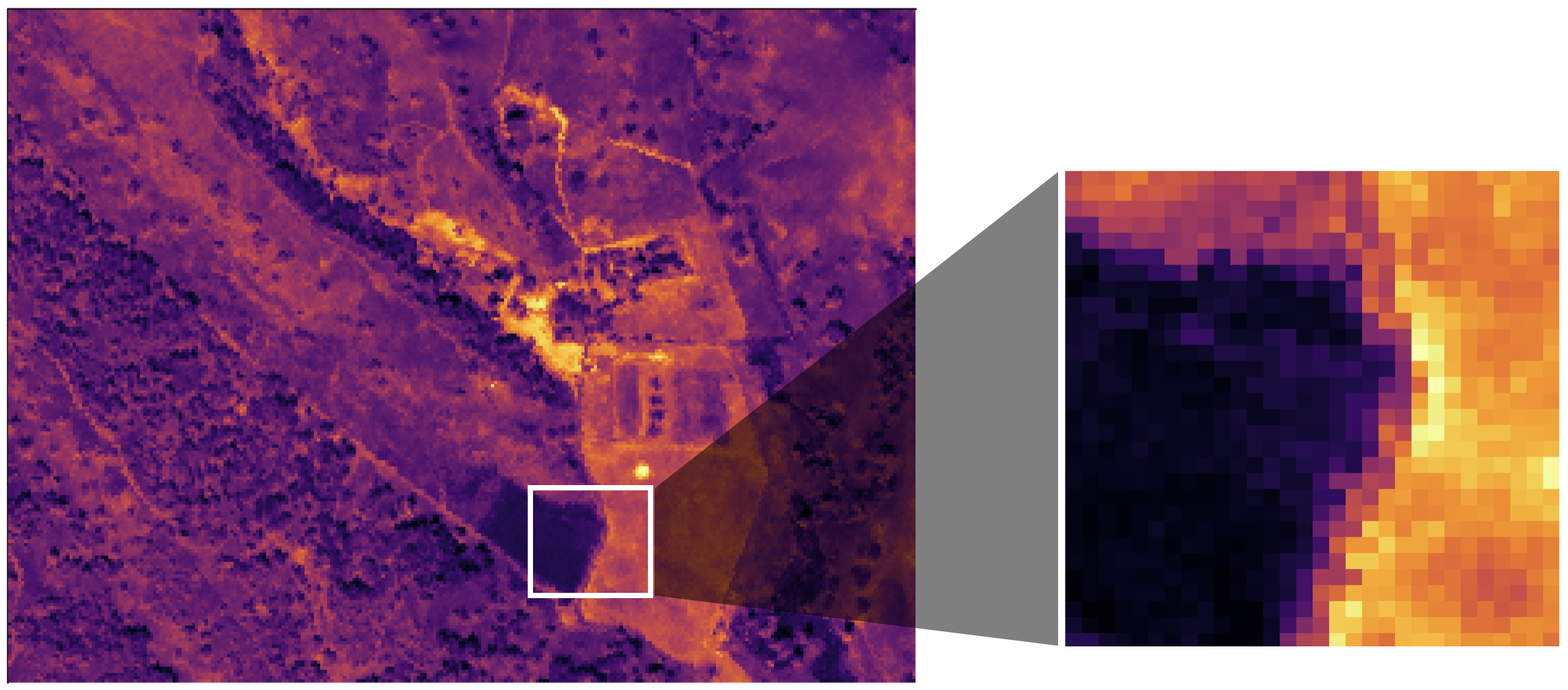}
    \caption{VSWIR Example. One band of the $30\times 30$ scene of Sedgwick Reserve, CA}
    \label{fig:sedgwick}
\end{figure}

We evaluate on a $30 \times 30$ subset from Sedgwick Reserve (approximately $34^\circ$N $120^\circ$W), Santa Barbara County, CA, covered by SHIFT in Spring 2022. Figure~\ref{fig:sedgwick} shows one radiance band. The scene contains three spectrally distinct subregions: coyote bush (left, dark purple), grassland (right, yellow), and a dirt road separating them, a natural testbed for evaluating spatial contrast preservation across time.


\subsubsection{Performance on Simulated Dataset}

We apply the filter with autoregressive weight $\alpha = 0.8$ and covariance inflation factor $\beta = 2$, and evaluate performacne across time, spectrum, and space.

\begin{figure}
    \centering
    \includegraphics[width=0.5\linewidth]{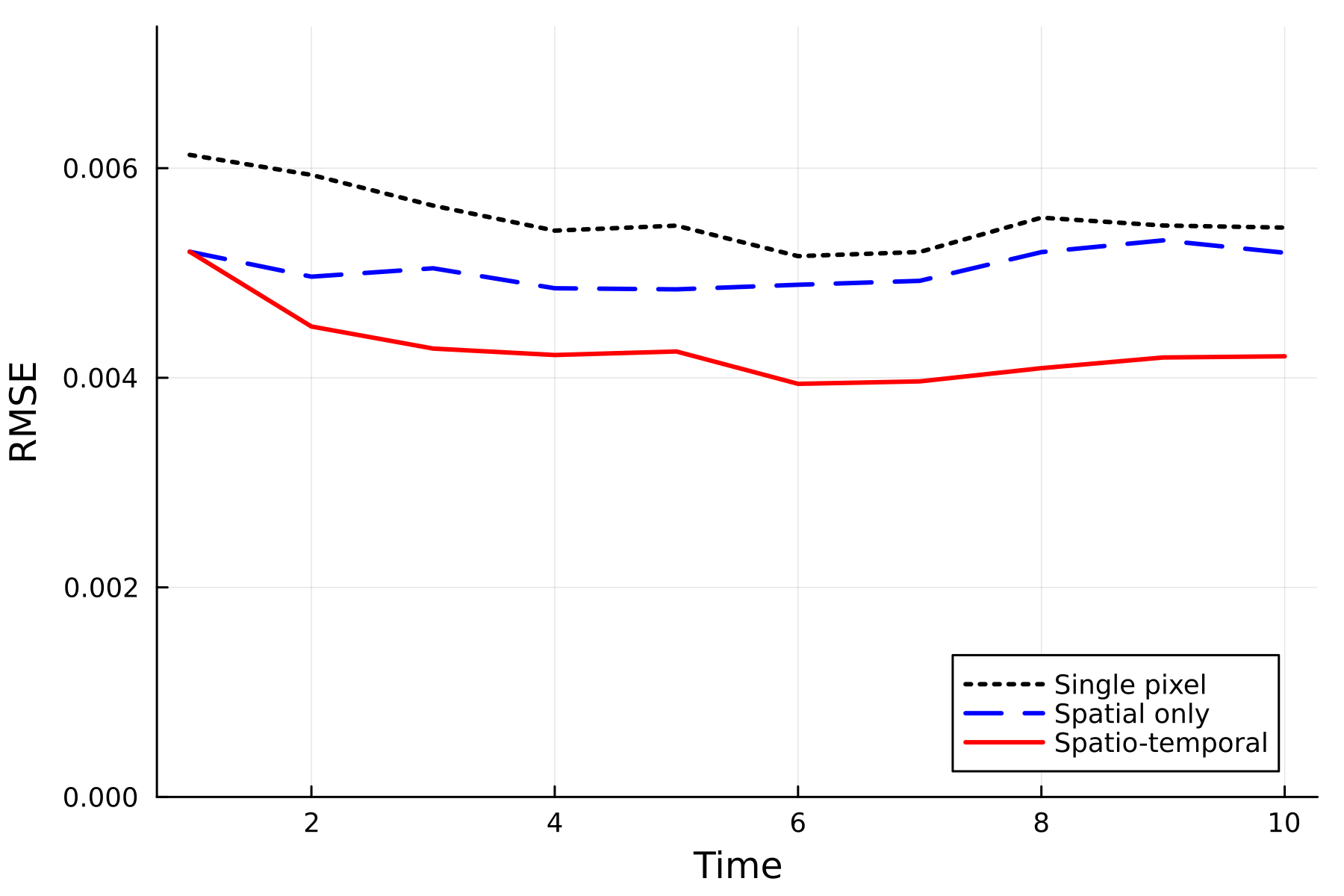}
    \caption{VSWIR Example. Overall RMSE as a function of time}
    \label{fig:errorOverTime}
\end{figure}

Figure~\ref{fig:errorOverTime} shows RMSE over time, averaged across wavelengths and spatial locations. Even though spatial correlations are modeled only in the atmosphere, the spatial retrieval improves surface estimates by constraining atmospheric estimates across pixels. The spatio-temporal algorithm further reduces RMSE by updating the surface prior from previous retrievals. At $t=1$, the two methods are identical since no past information is available. 
By $t=3$, the benefit of temporal coupling is clear, and the spatio-temporal method maintains lower error for all subsequent time steps.



\subsubsection*{Error across spectral bands}

\begin{figure}
\centering
\begin{subfigure}{.5\textwidth}
  \centering
  \includegraphics[width=0.9\linewidth]{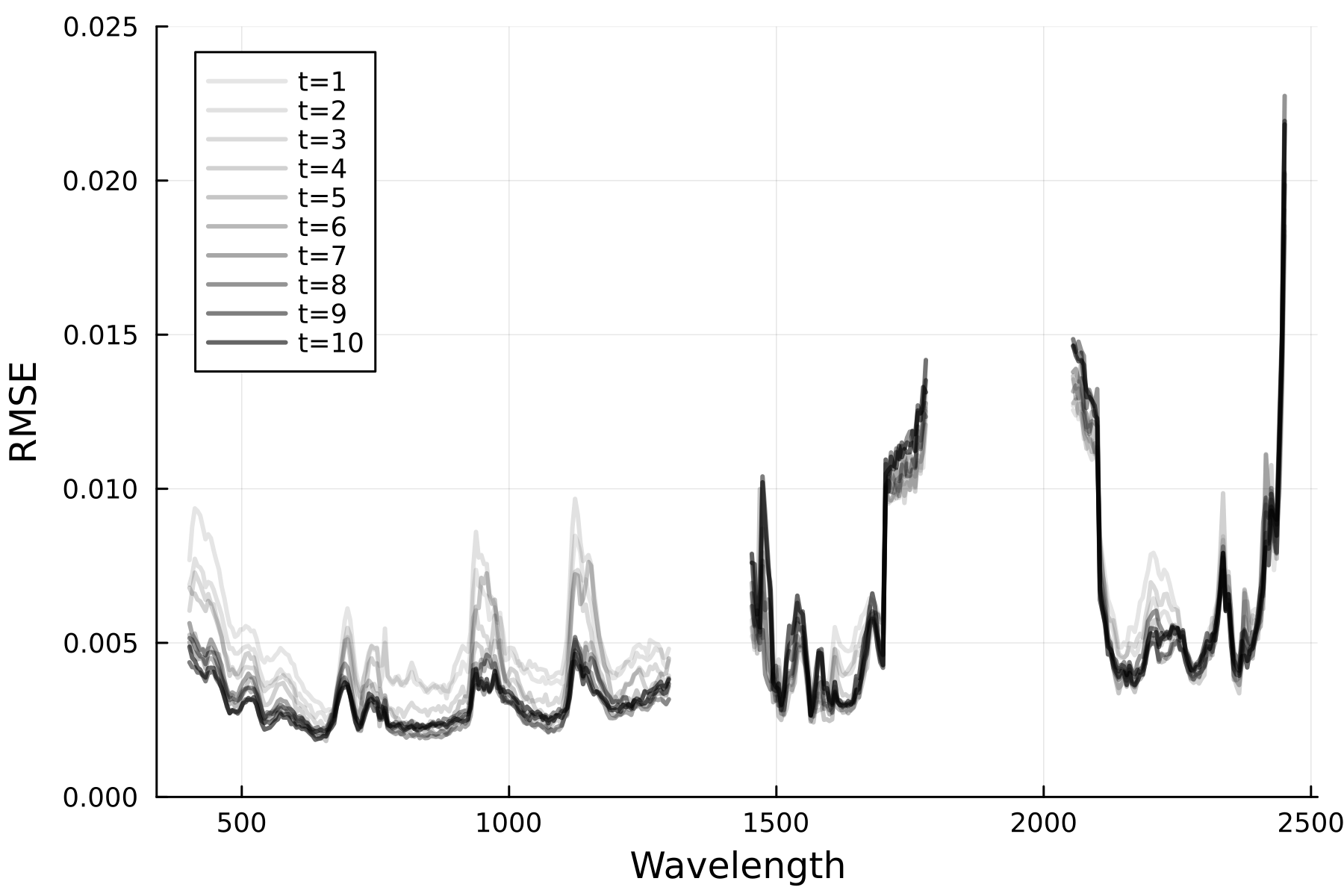}
  \caption{Single pixel retrieval}
  \label{fig:errorOverSpectra_orig}
\end{subfigure}%
\begin{subfigure}{.5\textwidth}
  \centering
  \includegraphics[width=0.9\linewidth]{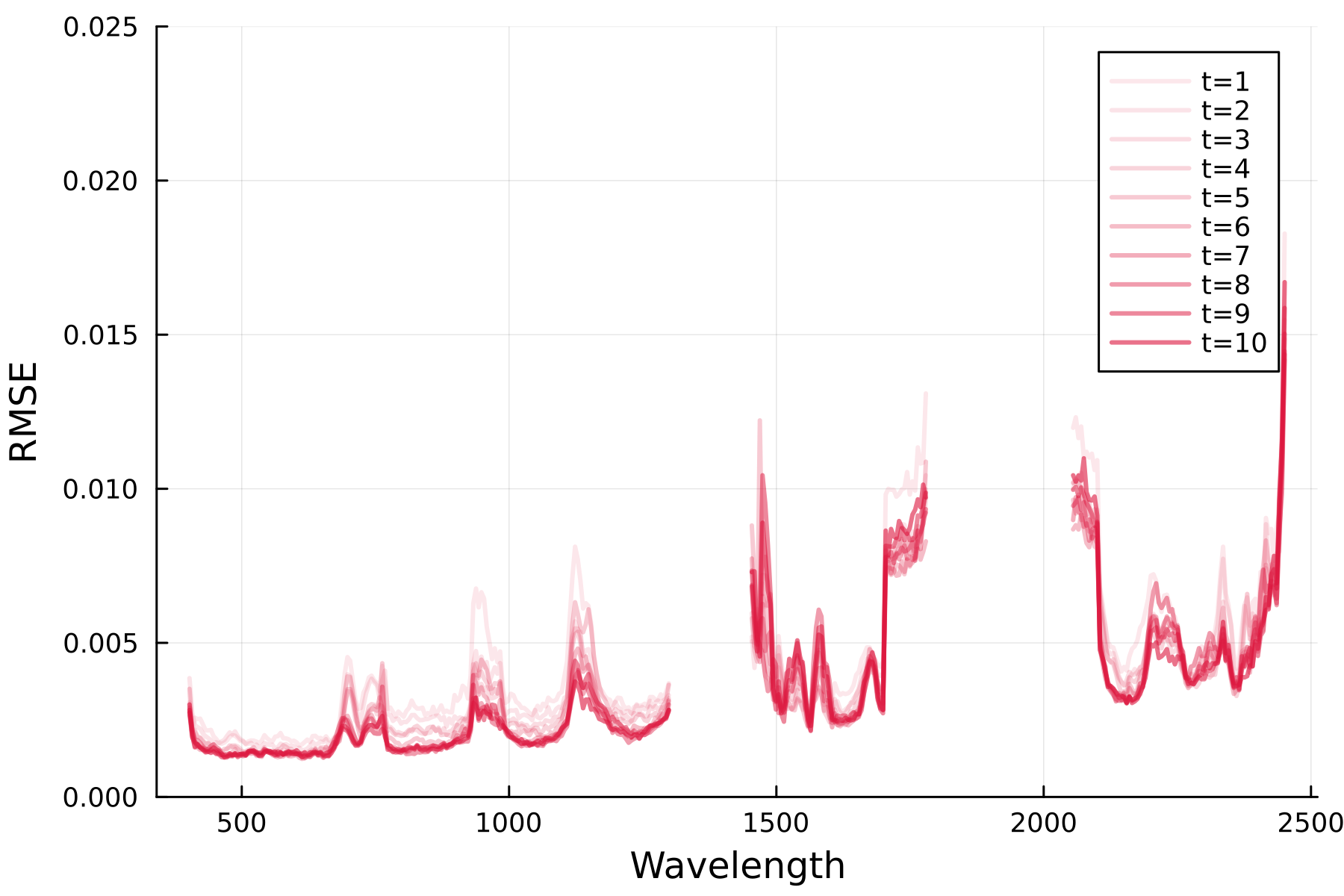}
  \caption{Spatiotemporal retrieval}
  \label{fig:errorOverSpectra_st}
\end{subfigure}
\caption{VSWIR Example. RMSE across the spectrum at each timestep.}
\label{fig:errorOverSpectra}
\end{figure}

Figure~\ref{fig:errorOverSpectra} shows RMSE per spectral band averaged over all spatial locations. The spatio-temporal method consistently reduces error across most of the spectrum, with the strongest gains in the 400–1400 nm range. Bands with strong water vapor absorption (1300–1450 nm and 1800–2100 nm) are excluded from the retrieval.

\begin{figure}
    \centering
    \includegraphics[width=0.45\linewidth]{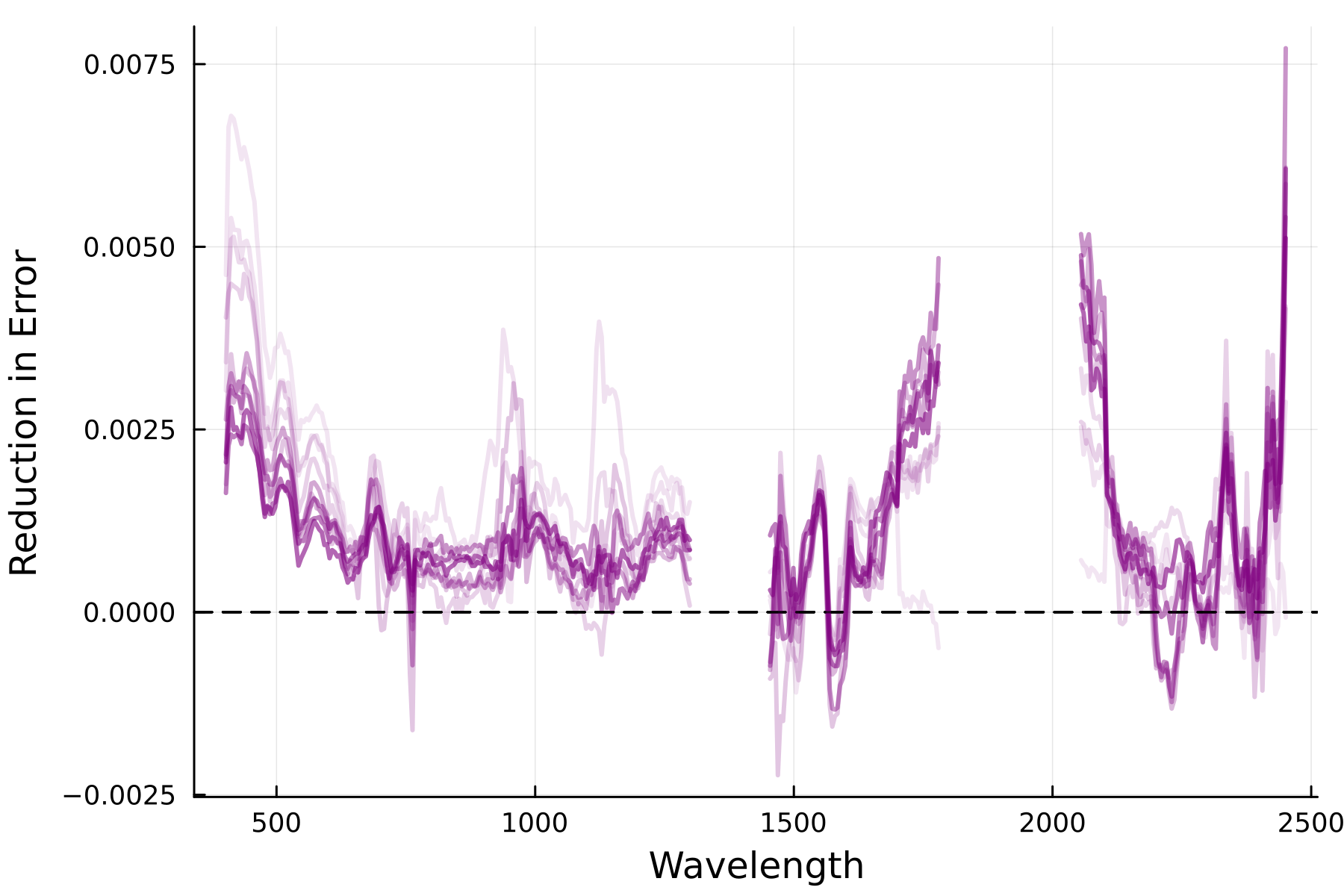}
    \caption{VSWIR Example. Reduction of spatio-temporal over single pixel retrieval, plotted over spectrum (positive values represent lower error in our spatio-temporal method).}
    \label{fig:errorOverSpectra_diff}
\end{figure}

Figure~\ref{fig:errorOverSpectra_diff} shows the RMSE difference across wavelengths, where positive values indicate spatio-temporal improvement. Most time steps show net improvement, with minor localized exceptions near 1500 nm and above 2000 nm. These small increases reflect how prior modification redistributes error across correlated spectral bands, but the overall trend is a consistent reduction when averaged across the spectrum.

\subsubsection*{Error across space}

\begin{figure}
    \centering
    \includegraphics[width=\linewidth]{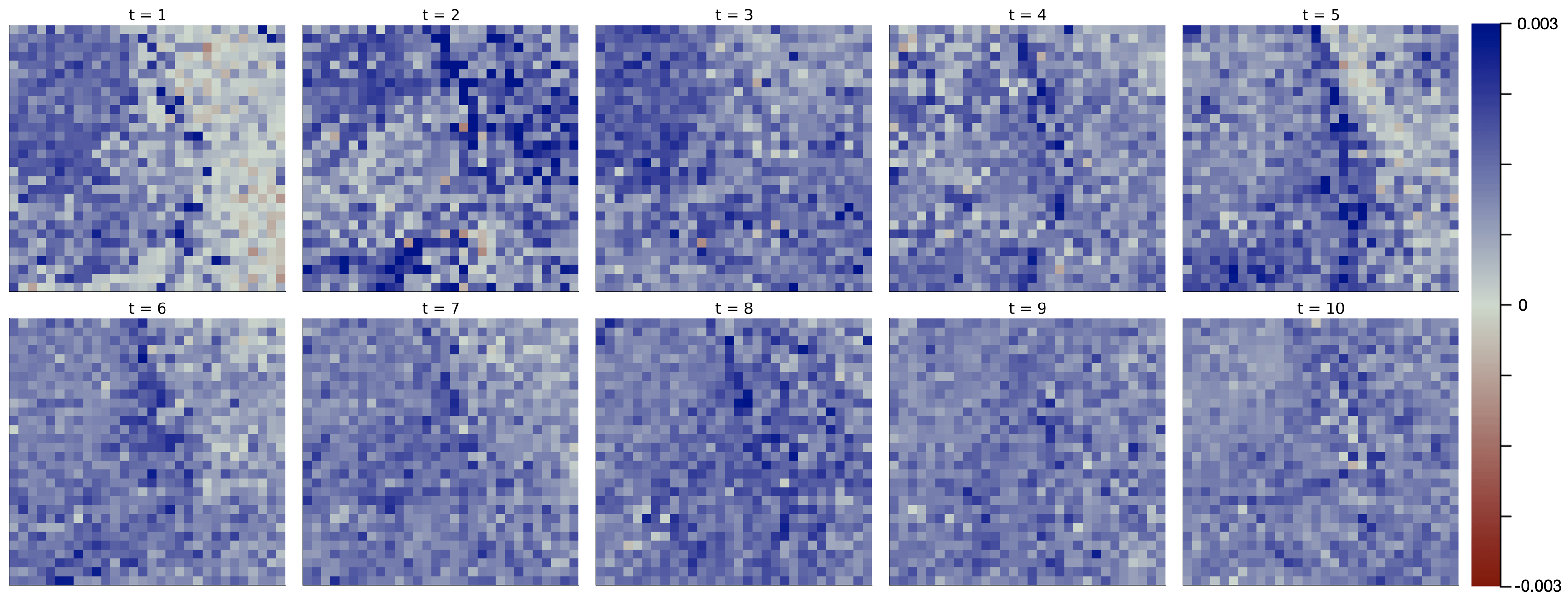}
    \caption{VSWIR Example. Reduction in relative error from the spatio-temporal method over space. Positive numbers, represented in blue, indicate lower error in our spatio-temporal method.}
    \label{fig:improvementHeatmap}
\end{figure}

To evaluate how retrieval performance varies spatially, we visualize the improvement in relative error of the spatio-temporal method over the single-pixel baseline across the full $30 \times 30$ spatial field in Figure~\ref{fig:improvementHeatmap}. 
Each heatmap shows the difference in error at a specific time step, with blue indicating improved accuracy from the spatio-temporal method and red indicating regions where it underperforms.

Improvements are broadly distributed and mostly positive, suggesting consistent error reduction across the scene. However, the strength of improvement varies both in space and time. 
At $t=1$, there is a clear difference between the two main subregions: the coyote bush (left) exhibits stronger improvements than the grassland (right), indicating that the spatial prior alone is more effective in the bush subregion at initialization.
Starting at $t=2$, as the filter begins propagating information temporally, improvements become more evenly distributed across the entire scene. This temporal smoothing effect gradually balances out the initial disparity in performance between surface types.


A few localized patches show minor degradation (red), but these tend to be small and isolated, likely due to local mismatches between the prior and the synthetic truth. Notably, the strongest and most consistent improvements are found in the narrow band corresponding to the dirt road between the two main subregions, visible especially at time steps 1, 2, 4, 5, 6, and 8. While the cause is not immediately obvious from the surface alone, this pattern is explained by the atmospheric retrieval behavior discussed in the following section.


\subsubsection*{Retrieval of atmospheric parameters}

Although surface reflectance is the primary objective, the algorithm's treatment of atmospheric parameters, particularly AOD, plays a critical supporting role. Despite being nuisance parameters, their influence on the nonlinear forward model makes them central to retrieval performance.


In nonlinear inverse problems like this one, multiple combinations of surface and atmospheric states can produce similar modeled radiances. Because most of the nonlinearity in the forward model originates from the atmosphere, and there are only two atmospheric variables, the optimization can exploit this flexibility by adjusting atmospheric parameters in unrealistic ways to achieve a better radiance fit. This trade-off leads to physically implausible atmospheric states and degraded surface retrievals.

\begin{figure}
\centering
\begin{subfigure}{\textwidth}
  \centering
  \includegraphics[width=0.8\linewidth]{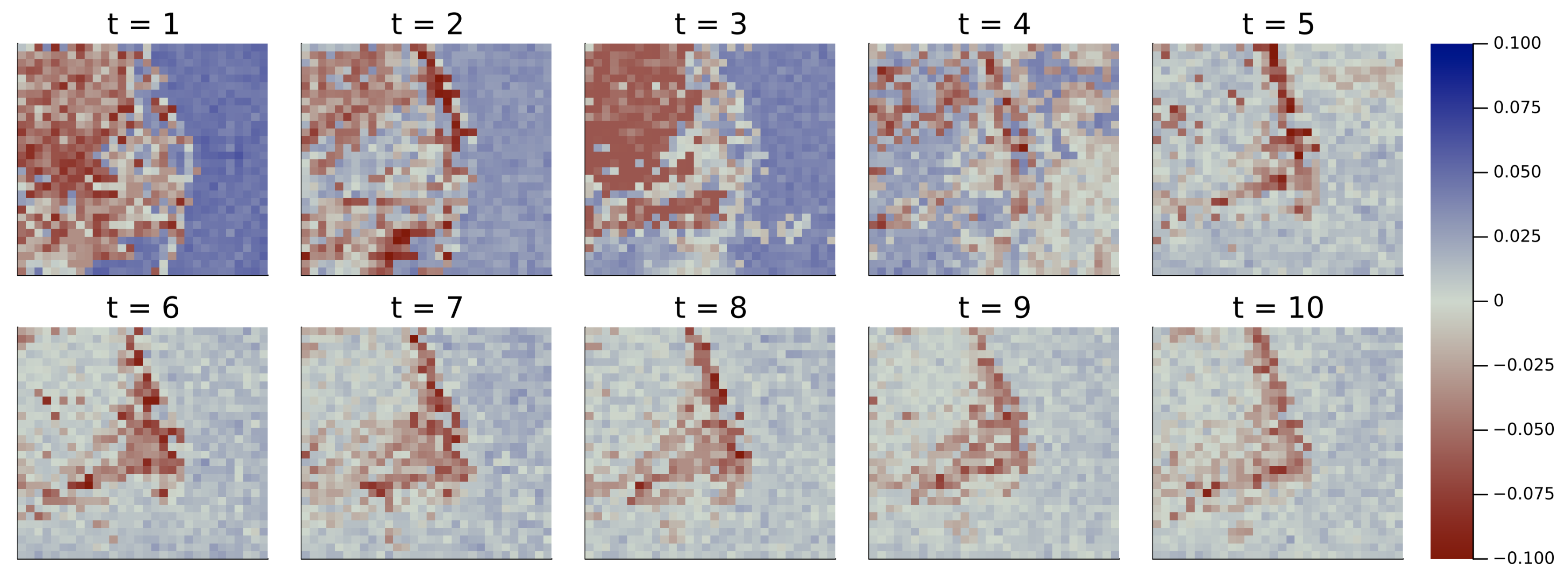}
  \caption{Single pixel retrieval}
  \label{fig:retrievedAOD_og}
   \par\medskip 
  \includegraphics[width=0.8\linewidth]{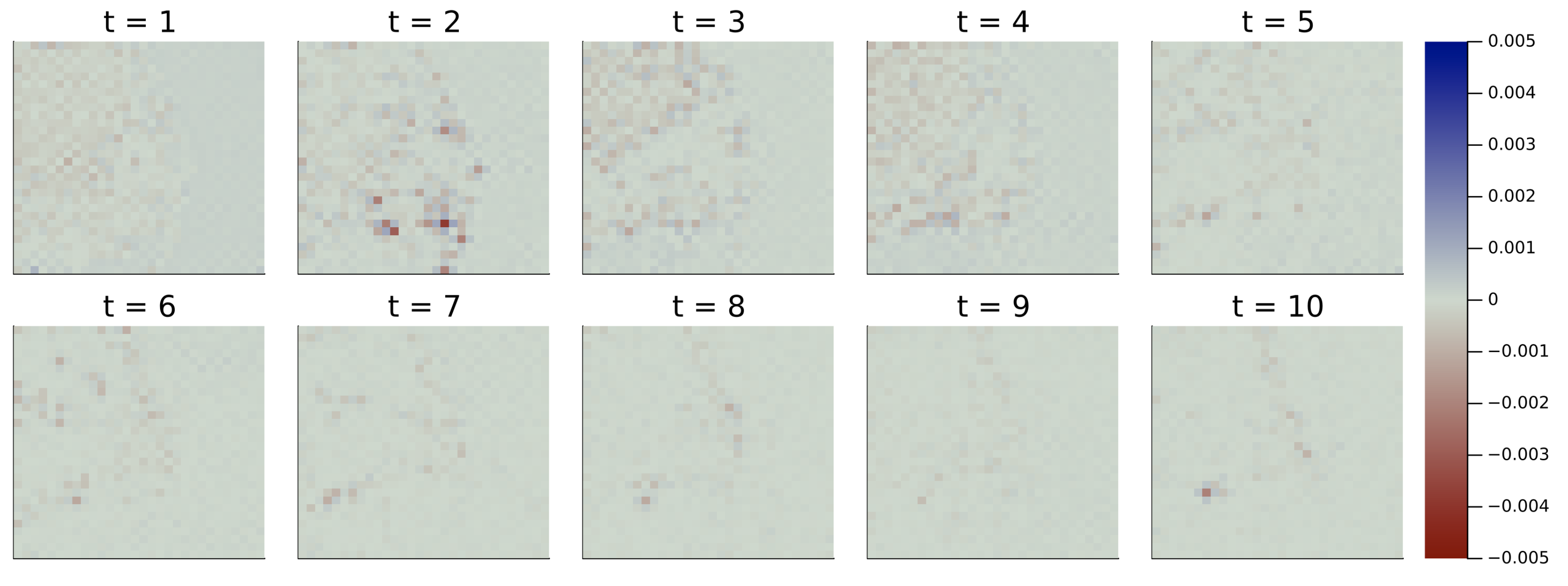}
  \caption{Spatio-temporal retrieval}
  \label{fig:retrievedAOD_st}
\end{subfigure}%
\caption{VSWIR Example. Retrieved aerosol optical depth (AOD) across the spatial domain, shown as deviations from the field mean.}
\label{fig:retrievedAOD}
\end{figure}

This phenomenon is evident in Figure~\ref{fig:retrievedAOD_og}, which shows retrieved AOD over time using the single-pixel method as deviations from the field mean. In many frames, especially at $t=1$ and $t=3$, retrieved AOD values cluster near the lower bounds of the retrieval domain. Moreover, these fields display structured spatial patterns that resemble surface features rather than realistic atmospheric variability – a clear sign that the optimizer is leveraging the atmosphere to compensate for errors elsewhere in the state. This also explains the road-shaped improvement pattern observed in Figure~\ref{fig:improvementHeatmap}: the road, being spectrally distinct from surrounding vegetation, induces particularly large atmospheric compensation in the single-pixel method, and the spatial prior effectively suppresses this behavior.



In contrast, the spatio-temporal method (Figure~\ref{fig:retrievedAOD_st}) enforces spatial consistency through the correlated atmospheric prior, preventing individual pixels from diverging to implausible atmospheric values. The resulting AOD fields are smoother and more physically plausible, with values generally remaining within 10\% of the spatial mean. Surface-like artifacts are substantially reduced compared to the single-pixel case. By regularizing the atmosphere in this way, the algorithm avoids sacrificing physical realism for numerical fit, and surface retrievals benefit as a result.

This outcome highlights the value of modeling spatial correlations in the atmosphere even when it is not the final product of interest. The atmospheric regularization acts as a structural safeguard that supports more stable and accurate surface inference, and is particularly impactful in scenes with high spectral contrast between surface types.

%

\section{Discussion}
\label{sec:discussion}

We presented a spatio-temporal filtering algorithm for remote sensing retrievals that exploits spatial and temporal correlations through a sparse precision matrix derived from conditional independence assumptions. The algorithm uses a spatially correlated atmospheric prior and a time-evolving surface prior, enabling improvements across retrieval problems where either component dominates. We demonstrated its flexibility and effectiveness through two representative use cases: NASA's OCO-2 mission, where the key quantity of interest is atmospheric CO$_2$, and VSWIR imaging spectroscopy, where the primary focus is surface reflectance.

The two experiments highlight how different components of the algorithm contribute depending on the application. In the OCO-2 case, performance gains were driven primarily by the spatial atmospheric prior, which captured meaningful cross-pixel correlations in atmospheric structure, yielding up to 50\% reduction in $XCO_2$ error over the single-pixel baseline. In the VSWIR experiment, both spatial and temporal components contributed: the nonlinearity of the forward model makes the spatial prior especially important, acting as a regularizer that discourages unphysical atmospheric solutions and improves surface estimates as a result \cite{leung_evaluating_2024}, with at least 20\% improvement in surface reflectance accuracy. Together, these experiments demonstrate the versatility of the approach.

As with any Bayesian method, performance is sensitive to prior and noise covariance configuration. An overly restrictive surface prior makes temporal filtering especially beneficial, while underestimating observation noise can cause over-regularization and degrade performance relative to independent retrievals. Careful tuning of these parameters, already standard practice in operational retrieval algorithms, remains important here.


Looking ahead, several directions remain for future work. Smoothing – conditioning on all observations to update prior time steps – could improve estimates at early time steps where filtering has not yet accumulated sufficient temporal information, as seen in Figure~\ref{fig:errorOverTime}. Moving beyond Gaussianity is also important, as the Gaussian assumption can break down under high humidity or aerosol loading \cite{leung_evaluating_2024}. Transport maps offer a promising pathway, with sparsity structures from graphical models informing their construction for scalable non-Gaussian inference \cite{marzouk_sampling_2016, spantini_coupling_2022}. These ideas have been extended to ensemble-based filtering and smoothing for sparse graphical models \cite{ramgraber_ensemble_2023, ramgraber_ensemble_2023-1, lunde_ensemble_2025}, and \cite{katzfuss_scalable_2023} demonstrates a related approach for non-Gaussian spatial models via sparse transport maps. Together, these tools offer a pathway for generalizing our approach beyond Gaussian assumptions while maintaining scalability.

\appendix
\section{Derivations}

\subsection{Applying conditional independence relations to the filtering distribution}
\label{sec:filterderivation}
\begin{align}
    p(\mathbf{x}_{1:m}^t &| \mathbf{y}_{1:m}^{1:t}) = p(\mathbf{s}_{1:m}^t, \mathbf{a}_{1:m}^t | \mathbf{y}_{1:m}^{1:t}) \nonumber\\
    &\propto p(\mathbf{y}_{1:m}^{t} | \mathbf{s}_{1:m}^t, \mathbf{a}_{1:m}^t) \, p(\mathbf{s}_{1:m}^t, \mathbf{a}_{1:m}^t | \mathbf{y}_{1:m}^{1:t-1})  \nonumber\\
    &= \left( \prod_{j=1}^m p(\mathbf{y}_{j}^{t} | \mathbf{s}_{j}^t, \mathbf{a}_{j}^t) \right) \, p(\mathbf{s}_{1:m}^t | \mathbf{y}_{1:m}^{1:t-1}) \,p( \mathbf{a}_{1:m}^t)  \nonumber\\
    &= \left( \prod_{j=1}^m p(\mathbf{y}_{j}^{t} | \mathbf{s}_{j}^t, \mathbf{a}_{j}^t) \right) \, \left[ \int p(\mathbf{s}_{1:m}^t, \mathbf{s}_{1:m}^{t-1} | \mathbf{y}_{1:m}^{1:t-1}) \,d\mathbf{s}_{1:m}^{t-1} \right] \,p( \mathbf{a}_{1:m}^t)  \nonumber\\
    &\propto \left( \prod_{j=1}^m p(\mathbf{y}_j^t | \mathbf{s}_j^t, \mathbf{a}_j^t)\right) \left[ \int \left( \prod_{j=1}^m p(\mathbf{s}_j^t | \mathbf{s}_j^{t-1}) \right) \,p(\mathbf{s}_{1:m}^{t-1} | \mathbf{y}_{1:m}^{1:t-1}) \, d\mathbf{s}_{1:m}^{t-1}  \right] \, p(\mathbf{a}_{1:m}^t)  \nonumber
\end{align}

\subsection{Expression for the cost minimization in MAP estimation}
\label{sec:simplifymapcost}
\begin{align}
    \mu^{t|t} \approx \mathbf{x}_{1:m}^* = \text{argmin}_{\mathbf{x}_{1:m}}\,\, -\log p(\mathbf{x}_{1:m}^t | \mathbf{y}_{1:m}^{1:t})  \nonumber
    = \text{argmin}_{\mathbf{x}_{1:m}} Q(\mathbf{x}_{1:m}^t; \mathbf{y}_{1:m}^t)  \nonumber
\end{align}
\begin{align}
    Q(\mathbf{x}_{1:m}; \mathbf{y}_{1:m}) &= \sum_{j = 1}^m ( \mathbf{y}_{j} - f(\mathbf{x}_j) )^\top \Gamma_\epsilon^{-1} ( \mathbf{y}_{j} - f(\mathbf{x}_j) ) + (\mathbf{x}_{1:m} - \mu)^\top \pmb\Lambda \, (\mathbf{x}_{1:m} - \mu)   \nonumber\\
     &=  \sum_{j = 1}^m  \left( \| \mathbf{y}_{j} - f(\mathbf{s}_j, \mathbf{a}_j) \|^2_{\Gamma_\epsilon} + \|\mathbf{s}_j - \mu_s \|^2_{\Gamma_s} \right) +  \sum_{i=1}^{n_a} \|\mathbf{a}_{i,1:m} - \mu_{a_i} \|^2_{\Sigma_{a_i}}  \nonumber
\end{align}

\section{Error in the matrix inversion}
\label{sec:matrixinverror}

\begin{figure}[H]
\centering
\begin{subfigure}{.45\textwidth}
  \centering
  \includegraphics[width=0.9\linewidth]{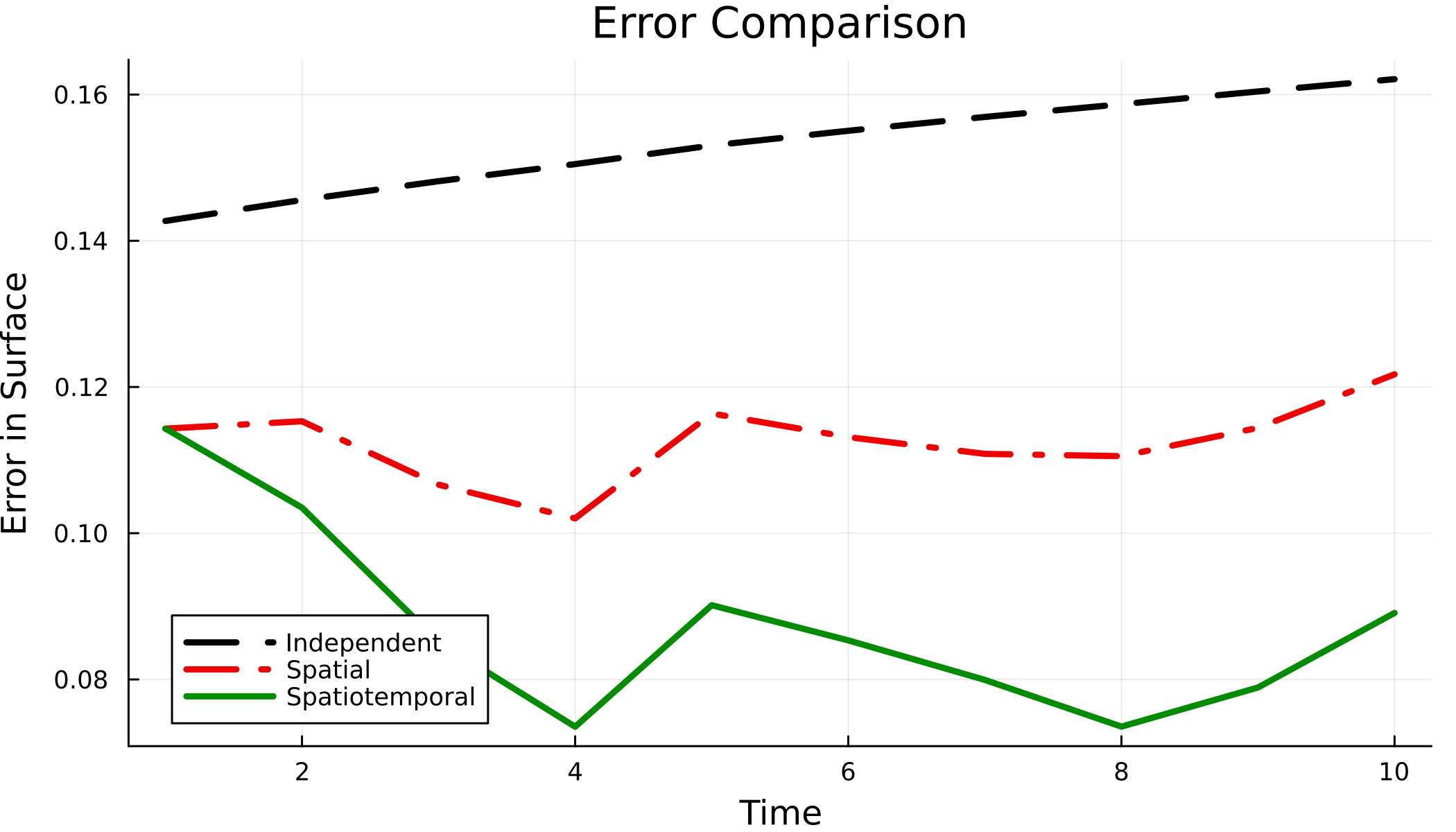}
  \caption{Surface retrieval error over time using the approximate marginal covariance.}
  \label{fig:err_over_time_toy}
\end{subfigure}%
\hspace{0.03\textwidth}
\begin{subfigure}{.5\textwidth}
  \centering
  \includegraphics[width=0.9\linewidth]{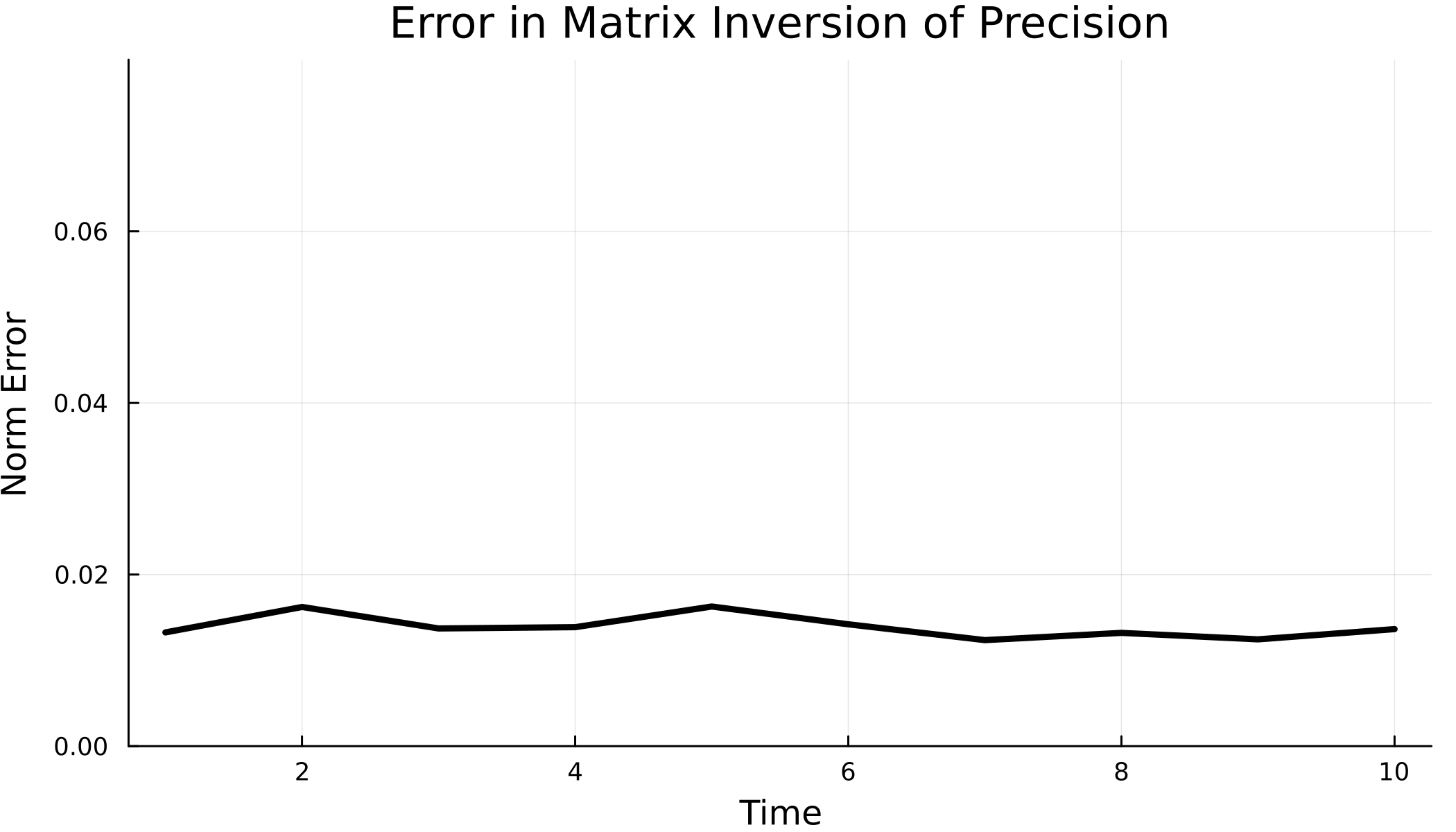}
  \caption{Norm error of the approximate marginal covariance relative to the exact inverse of the full precision.}
  \label{fig:error_matrixnorm}
\end{subfigure}
\caption{Error in surface retrieval and in marginal covariance over time for a synthetic problem with $n_s=3$ and $m=100$.}
\label{fig:matrixinverror}
\end{figure}

The error in the marginal covariance using the approximate block inversion method presented in Section~\ref{sec:approximatemarginals} is empirically assessed on a synthetic spatio-temporal problem with $n_s=3$ and $m=100$.
Figure~\ref{fig:matrixinverror} reports both the resulting surface retrieval error over time and the norm error of the approximate marginal covariance relative to the exact inverse of the full precision matrix.

\bibliographystyle{siamplain}
\bibliography{references}

@incollection{marzouk_sampling_2016,
	address = {Cham},
	title = {Sampling via {Measure} {Transport}: {An} {Introduction}},
	isbn = {978-3-319-11259-6},
	booktitle = {Handbook of {Uncertainty} {Quantification}},
	publisher = {Springer International Publishing},
	author = {Marzouk, Youssef and Moselhy, Tarek and Parno, Matthew and Spantini, Alessio},
	editor = {Ghanem, Roger and Higdon, David and Owhadi, Houman},
	year = {2016},
	doi = {10.1007/978-3-319-11259-6_23-1},
	pages = {1--41},
}

@article{lindgren_spde_2022,
	series = {Special {Issue}: {The} {Impact} of {Spatial} {Statistics}},
	title = {The {SPDE} approach for {Gaussian} and non-{Gaussian} fields: 10 years and still running},
	volume = {50},
	issn = {2211-6753},
	shorttitle = {The {SPDE} approach for {Gaussian} and non-{Gaussian} fields},
	doi = {10.1016/j.spasta.2022.100599},
	journal = {Spatial Statistics},
	author = {Lindgren, Finn and Bolin, David and Rue, Håvard},
	month = aug,
	year = {2022},
	pages = {100599},
}

@article{lindgren_explicit_2011,
	title = {An explicit link between {Gaussian} fields and {Gaussian} {Markov} random fields: the stochastic partial differential equation approach},
	volume = {73},
	issn = {1467-9868},
	shorttitle = {An explicit link between {Gaussian} fields and {Gaussian} {Markov} random fields},
	doi = {10.1111/j.1467-9868.2011.00777.x},
	language = {en},
	number = {4},
	journal = {Journal of the Royal Statistical Society: Series B (Statistical Methodology)},
	author = {Lindgren, Finn and Rue, Håvard and Lindström, Johan},
	year = {2011},
	pages = {423--498},
}

@article{cawse-nicholson_nasas_2021,
	title = {{NASA}'s surface biology and geology designated observable: {A} perspective on surface imaging algorithms},
	volume = {257},
	issn = {0034-4257},
	shorttitle = {{NASA}'s surface biology and geology designated observable},
	doi = {10.1016/j.rse.2021.112349},
	journal = {Remote Sensing of Environment},
	author = {Cawse-Nicholson, Kerry and Townsend, Philip A. and Schimel, David and Assiri, Ali M. and Blake, Pamela L. and Buongiorno, Maria Fabrizia and Campbell, Petya and Carmon, Nimrod and Casey, Kimberly A. and Correa-Pabón, Rosa Elvira and Dahlin, Kyla M. and Dashti, Hamid and Dennison, Philip E. and Dierssen, Heidi and Erickson, Adam and Fisher, Joshua B. and Frouin, Robert and Gatebe, Charles K. and Gholizadeh, Hamed and Gierach, Michelle and Glenn, Nancy F. and Goodman, James A. and Griffith, Daniel M. and Guild, Liane and Hakkenberg, Christopher R. and Hochberg, Eric J. and Holmes, Thomas R. H. and Hu, Chuanmin and Hulley, Glynn and Huemmrich, Karl F. and Kudela, Raphael M. and Kokaly, Raymond F. and Lee, Christine M. and Martin, Roberta and Miller, Charles E. and Moses, Wesley J. and Muller-Karger, Frank E. and Ortiz, Joseph D. and Otis, Daniel B. and Pahlevan, Nima and Painter, Thomas H. and Pavlick, Ryan and Poulter, Ben and Qi, Yi and Realmuto, Vincent J. and Roberts, Dar and Schaepman, Michael E. and Schneider, Fabian D. and Schwandner, Florian M. and Serbin, Shawn P. and Shiklomanov, Alexey N. and Stavros, E. Natasha and Thompson, David R. and Torres-Perez, Juan L. and Turpie, Kevin R. and Tzortziou, Maria and Ustin, Susan and Yu, Qian and Yusup, Yusri and Zhang, Qingyuan},
	month = may,
	year = {2021},
	pages = {112349},
}

@article{cawse-nicholson_surface_2023,
	title = {Surface {Biology} and {Geology} imaging spectrometer: {A} case study to optimize the mission design using intrinsic dimensionality},
	volume = {290},
	issn = {0034-4257},
	shorttitle = {Surface {Biology} and {Geology} imaging spectrometer},
	doi = {10.1016/j.rse.2023.113534},
	journal = {Remote Sensing of Environment},
	author = {Cawse-Nicholson, K. and Raiho, A. M. and Thompson, D. R. and Hulley, G. C. and Miller, C. E. and Miner, K. R. and Poulter, B. and Schimel, D. and Schneider, F. D. and Townsend, P. A. and Zareh, S. K.},
	month = may,
	year = {2023},
	pages = {113534},
}

@article{zilber_spatial_2022,
	title = {Spatial {Surface} {Reflectance} {Retrievals} for {Visible}/{Shortwave} {Infrared} {Remote} {Sensing} via {Gaussian} {Process} {Priors}},
	volume = {14},
	copyright = {http://creativecommons.org/licenses/by/3.0/},
	issn = {2072-4292},
	doi = {10.3390/rs14092183},
	language = {en},
	number = {9},
	journal = {Remote Sensing},
	author = {Zilber, Daniel and Thompson, David R. and Katzfuss, Matthias and Natraj, Vijay and Hobbs, Jonathan and Braverman, Amy},
	month = jan,
	year = {2022},
	note = {Number: 9
Publisher: Multidisciplinary Digital Publishing Institute},
	pages = {2183},
}

@article{thompson_optimal_2018,
	title = {Optimal estimation for imaging spectrometer atmospheric correction},
	volume = {216},
	issn = {0034-4257},
	doi = {10.1016/j.rse.2018.07.003},
	journal = {Remote Sensing of Environment},
	author = {Thompson, David R. and Natraj, Vijay and Green, Robert O. and Helmlinger, Mark C. and Gao, Bo-Cai and Eastwood, Michael L.},
	month = oct,
	year = {2018},
	pages = {355--373},
}

@article{hobbs_spatial_2021,
	title = {Spatial {Retrievals} of {Atmospheric} {Carbon} {Dioxide} from {Satellite} {Observations}},
	volume = {13},
	copyright = {http://creativecommons.org/licenses/by/3.0/},
	issn = {2072-4292},
	doi = {10.3390/rs13040571},
	language = {en},
	number = {4},
	journal = {Remote Sensing},
	author = {Hobbs, Jonathan and Katzfuss, Matthias and Zilber, Daniel and Brynjarsdóttir, Jenný and Mondal, Anirban and Berrocal, Veronica},
	month = jan,
	year = {2021},
	note = {Number: 4
Publisher: Multidisciplinary Digital Publishing Institute},
	pages = {571},
}

@inproceedings{hamlin_imaging_2011,
	title = {Imaging spectrometer science measurements for {Terrestrial} {Ecology}: {AVIRIS} and new developments},
	shorttitle = {Imaging spectrometer science measurements for {Terrestrial} {Ecology}},
	doi = {10.1109/AERO.2011.5747395},
	booktitle = {2011 {Aerospace} {Conference}},
	author = {Hamlin, L. and Green, R. O. and Mouroulis, P. and Eastwood, M. and Wilson, D. and Dudik, M. and Paine, C.},
	month = mar,
	year = {2011},
	note = {ISSN: 1095-323X},
	pages = {1--7},
}

@book{rasmussen_gaussian_2006,
	address = {Cambridge, Mass},
	series = {Adaptive computation and machine learning},
	title = {Gaussian processes for machine learning},
	isbn = {978-0-262-18253-9},
	language = {en},
	publisher = {MIT Press},
	author = {Rasmussen, Carl Edward and Williams, Christopher K. I.},
	year = {2006},
	note = {OCLC: ocm61285753},
}

@book{wikle_spatio-temporal_2019,
	address = {Boca Raton},
	series = {Chapman \& {Hall}/{CRC} the {R} series},
	title = {Spatio-temporal statistics with {R}},
	isbn = {978-1-351-76972-3 978-0-429-64978-3},
	language = {en},
	publisher = {CRC Press, Taylor and Francis Group},
	author = {Wikle, Christopher K. and Zammit-Mangion, Andrew and Cressie, Noel A. C.},
	year = {2019},
}

@article{leung_evaluating_2024,
	title = {Evaluating the Accuracy of Gaussian Approximations in {VSWIR} Imaging Spectroscopy Retrievals},
	volume = {62},
	issn = {1558-0644},
	doi = {10.1109/TGRS.2024.3411916},
	pages = {1--12},
	journaltitle = {{IEEE} Transactions on Geoscience and Remote Sensing},
	author = {Leung, Kelvin M. and Thompson, David R. and Susiluoto, Jouni and Jagalur, Jayanth and Braverman, Amy and Marzouk, Youssef},
	date = {2024},
}

@misc{green_2023_aviris,
  author       = {Green, R. O. and Brodrick, P. G. and Chapman, J. W. and Eastwood, M. and Geier, S. and Helmlinger, M. and Lundeen, S. R. and Olson-Duvall, W. and Pavlick, R. and Rios, L. M. and Thompson, D. R. and Thorpe, A. K.},
  title        = {{AVIRIS-NG L1B Calibrated Radiance, Facility Instrument Collection, V1}},
  year         = {2023},
  publisher    = {ORNL DAAC},
  location     = {Oak Ridge, Tennessee, USA},
  url          = {https://doi.org/10.3334/ORNLDAAC/2095},
  doi          = {10.3334/ORNLDAAC/2095}
}

@article{cressie_fixed_2008,
	title = {Fixed rank kriging for very large spatial data sets},
	volume = {70},
	issn = {1467-9868},
	doi = {10.1111/j.1467-9868.2007.00633.x},
	language = {en},
	number = {1},
	journal = {Journal of the Royal Statistical Society: Series B (Statistical Methodology)},
	author = {Cressie, Noel and Johannesson, Gardar},
	year = {2008},
	pages = {209--226},
}

@article{banerjee_gaussian_2008,
	title = {Gaussian {Predictive} {Process} {Models} for {Large} {Spatial} {Data} {Sets}},
	volume = {70},
	issn = {1369-7412},
	number = {4},
	journal = {Journal of the Royal Statistical Society. Series B (Statistical Methodology)},
	author = {Banerjee, Sudipto and Gelfand, Alan E. and Finley, Andrew O. and Sang, Huiyan},
	year = {2008},
	note = {Publisher: [Royal Statistical Society, Wiley]},
	pages = {825--848},
}

@article{wainwright_graphical_2008,
	title = {Graphical {Models}, {Exponential} {Families}, and {Variational} {Inference}},
	volume = {1},
	issn = {1935-8237, 1935-8245},
	doi = {10.1561/2200000001},
	language = {English},
	number = {1–2},
	journal = {Foundations and Trends® in Machine Learning},
	author = {Wainwright, Martin J. and Jordan, Michael I.},
	month = nov,
	year = {2008},
	note = {Publisher: Now Publishers, Inc.},
	pages = {1--305},
}

@article{heaton_case_2019,
	title = {A {Case} {Study} {Competition} {Among} {Methods} for {Analyzing} {Large} {Spatial} {Data}},
	volume = {24},
	issn = {1537-2693},
	doi = {10.1007/s13253-018-00348-w},
	language = {en},
	number = {3},
	journal = {Journal of Agricultural, Biological and Environmental Statistics},
	author = {Heaton, Matthew J. and Datta, Abhirup and Finley, Andrew O. and Furrer, Reinhard and Guinness, Joseph and Guhaniyogi, Rajarshi and Gerber, Florian and Gramacy, Robert B. and Hammerling, Dorit and Katzfuss, Matthias and Lindgren, Finn and Nychka, Douglas W. and Sun, Furong and Zammit-Mangion, Andrew},
	month = sep,
	year = {2019},
	pages = {398--425},
}

@article{finley_improving_2009,
	title = {Improving the performance of predictive process modeling for large datasets},
	volume = {53},
	issn = {0167-9473},
	doi = {10.1016/j.csda.2008.09.008},
	number = {8},
	journal = {Computational Statistics \& Data Analysis},
	author = {Finley, Andrew O. and Sang, Huiyan and Banerjee, Sudipto and Gelfand, Alan E.},
	month = jun,
	year = {2009},
	pages = {2873--2884},
}

@article{kaufman_covariance_2008,
	title = {Covariance {Tapering} for {Likelihood}-{Based} {Estimation} in {Large} {Spatial} {Data} {Sets}},
	volume = {103},
	issn = {0162-1459},
	doi = {10.1198/016214508000000959},
	number = {484},
	journal = {Journal of the American Statistical Association},
	author = {Kaufman, Cari G. and Schervish, Mark J. and Nychka, Douglas W.},
	month = dec,
	year = {2008},
	note = {Publisher: Taylor \& Francis},
	pages = {1545--1555},
}

@article{vecchia_estimation_1988,
	title = {Estimation and {Model} {Identification} for {Continuous} {Spatial} {Processes}},
	volume = {50},
	copyright = {© 1988 Royal Statistical Society},
	issn = {2517-6161},
	doi = {10.1111/j.2517-6161.1988.tb01729.x},
	language = {en},
	number = {2},
	journal = {Journal of the Royal Statistical Society: Series B (Methodological)},
	author = {Vecchia, A. V.},
	year = {1988},
	pages = {297--312},
}

@article{stein_approximating_2004,
	title = {Approximating {Likelihoods} for {Large} {Spatial} {Data} {Sets}},
	volume = {66},
	issn = {1369-7412},
	doi = {10.1046/j.1369-7412.2003.05512.x},
	number = {2},
	journal = {Journal of the Royal Statistical Society Series B: Statistical Methodology},
	author = {Stein, Michael L. and Chi, Zhiyi and Welty, Leah J.},
	month = may,
	year = {2004},
	pages = {275--296},
}

@article{katzfuss_multi-resolution_2017,
	title = {A {Multi}-{Resolution} {Approximation} for {Massive} {Spatial} {Datasets}},
	volume = {112},
	issn = {0162-1459},
	doi = {10.1080/01621459.2015.1123632},
	number = {517},
	journal = {Journal of the American Statistical Association},
	author = {Katzfuss, Matthias},
	month = jan,
	year = {2017},
	note = {Publisher: Taylor \& Francis},
	pages = {201--214},
}

@article{katzfuss_general_2021,
	title = {A {General} {Framework} for {Vecchia} {Approximations} of {Gaussian} {Processes}},
	volume = {36},
	issn = {0883-4237, 2168-8745},
	doi = {10.1214/19-STS755},
	number = {1},
	journal = {Statistical Science},
	author = {Katzfuss, Matthias and Guinness, Joseph},
	month = feb,
	year = {2021},
	note = {Publisher: Institute of Mathematical Statistics},
	pages = {124--141},
}

@article{katzfuss_scalable_2023,
	title = {Scalable {Bayesian} {Transport} {Maps} for {High}-{Dimensional} {Non}-{Gaussian} {Spatial} {Fields}},
	volume = {0},
	issn = {0162-1459},
	doi = {10.1080/01621459.2023.2197158},
	number = {0},
	journal = {Journal of the American Statistical Association},
	author = {Katzfuss, Matthias and Schäfer, Florian},
	year = {2023},
	note = {Publisher: Taylor \& Francis},
	pages = {1--15},
}

@article{furrer_covariance_2006,
	title = {Covariance {Tapering} for {Interpolation} of {Large} {Spatial} {Datasets}},
	volume = {15},
	issn = {1061-8600},
	number = {3},
	journal = {Journal of Computational and Graphical Statistics},
	author = {Furrer, Reinhard and Genton, Marc G. and Nychka, Douglas},
	year = {2006},
	note = {Publisher: [American Statistical Association, Taylor \& Francis, Ltd., Institute of Mathematical Statistics, Interface Foundation of America]},
	pages = {502--523},
}

@book{lauritzen_graphical_1996,
	address = {Oxford, New York},
	series = {Oxford {Statistical} {Science} {Series}},
	title = {Graphical {Models}},
	isbn = {978-0-19-852219-5},
	publisher = {Oxford University Press},
	author = {Lauritzen, Steffen L.},
	month = jul,
	year = {1996},
}

@article{jordan_graphical_2004,
	title = {Graphical {Models}},
	volume = {19},
	issn = {0883-4237, 2168-8745},
	doi = {10.1214/088342304000000026},
	number = {1},
	journal = {Statistical Science},
	author = {Jordan, Michael I.},
	month = feb,
	year = {2004},
	note = {Publisher: Institute of Mathematical Statistics},
	pages = {140--155},
}

@book{murphy_machine_2012,
	title = {Machine {Learning}: {A} {Probabilistic} {Perspective}},
	isbn = {978-0-262-01802-9},
	shorttitle = {Machine {Learning}},
	language = {en},
	publisher = {MIT Press},
	author = {Murphy, Kevin P.},
	month = aug,
	year = {2012},
}

@article{eckert_spatially_2023,
	title = {Spatially constrained retrieval for imaging spectroscopy},
	author = {Eckert, Regina and Mauceri, Steffen and Thompson, David and Fahlen, Jay and Brodrick, Phil},
	year = {2023},
}

@article{nychka_multiresolution_2015,
	title = {A {Multiresolution} {Gaussian} {Process} {Model} for the {Analysis} of {Large} {Spatial} {Datasets}},
	volume = {24},
	issn = {1061-8600},
	doi = {10.1080/10618600.2014.914946},
	number = {2},
	journal = {Journal of Computational and Graphical Statistics},
	author = {Nychka, Douglas and Bandyopadhyay, Soutir and Hammerling, Dorit and Lindgren, Finn and Sain, Stephan},
	month = apr,
	year = {2015},
	note = {Publisher: Taylor \& Francis},
	pages = {579--599},
}

@article{spantini_coupling_2022,
	title = {Coupling {Techniques} for {Nonlinear} {Ensemble} {Filtering}},
	volume = {64},
	issn = {0036-1445},
	doi = {10.1137/20M1312204},
	number = {4},
	journal = {SIAM Review},
	author = {Spantini, Alessio and Baptista, Ricardo and Marzouk, Youssef},
	month = nov,
	year = {2022},
	note = {Publisher: Society for Industrial and Applied Mathematics},
	pages = {921--953},
}

@article{houtekamer_data_1998,
	title = {Data {Assimilation} {Using} an {Ensemble} {Kalman} {Filter} {Technique}},
	volume = {126},
	issn = {1520-0493, 0027-0644},
	doi = {10.1175/1520-0493(1998)126<0796:DAUAEK>2.0.CO;2},
	language = {EN},
	number = {3},
	journal = {Monthly Weather Review},
	author = {Houtekamer, P. L. and Mitchell, Herschel L.},
	month = mar,
	year = {1998},
	note = {Publisher: American Meteorological Society
Section: Monthly Weather Review},
	pages = {796--811},
}

@article{evensen_ensemble_2003,
	title = {The {Ensemble} {Kalman} {Filter}: theoretical formulation and practical implementation},
	volume = {53},
	issn = {1616-7228},
	shorttitle = {The {Ensemble} {Kalman} {Filter}},
	doi = {10.1007/s10236-003-0036-9},
	language = {en},
	number = {4},
	journal = {Ocean Dynamics},
	author = {Evensen, Geir},
	month = nov,
	year = {2003},
	pages = {343--367},
}

@article{datta_nearest-neighbor_2016,
	title = {On nearest-neighbor {Gaussian} process models for massive spatial data},
	volume = {8},
	copyright = {© 2016 Wiley Periodicals, Inc.},
	issn = {1939-0068},
	doi = {10.1002/wics.1383},
	language = {en},
	number = {5},
	journal = {WIREs Computational Statistics},
	author = {Datta, Abhirup and Banerjee, Sudipto and Finley, Andrew O. and Gelfand, Alan E.},
	year = {2016},
	pages = {162--171},
}

@inproceedings{wilson_kernel_2015,
	address = {Lille, France},
	series = {{ICML}'15},
	title = {Kernel interpolation for scalable structured {Gaussian} processes ({KISS}-{GP})},
	booktitle = {Proceedings of the 32nd {International} {Conference} on {International} {Conference} on {Machine} {Learning} - {Volume} 37},
	publisher = {JMLR.org},
	author = {Wilson, Andrew Gordon and Nickisch, Hannes},
	month = jul,
	year = {2015},
	pages = {1775--1784},
}

@article{ramgraber_ensemble_2023,
	title = {Ensemble transport smoothing. {Part} {I}: {Unified} framework},
	volume = {17},
	issn = {2590-0552},
	shorttitle = {Ensemble transport smoothing. {Part} {I}},
	doi = {10.1016/j.jcpx.2023.100134},
	journal = {Journal of Computational Physics: X},
	author = {Ramgraber, Maximilian and Baptista, Ricardo and McLaughlin, Dennis and Marzouk, Youssef},
	month = nov,
	year = {2023},
	pages = {100134},
}

@article{anderson_optimal_1982,
	title = {Optimal {Filtering}},
	volume = {12},
	issn = {2168-2909},
	doi = {10.1109/TSMC.1982.4308806},
	number = {2},
	journal = {IEEE Transactions on Systems, Man, and Cybernetics},
	author = {Anderson, Brian D. O. and Moore, John B. and Eslami, Mansour},
	month = mar,
	year = {1982},
	pages = {235--236},
}

@article{singh_inverse_2023,
	title = {Inverse {Extended} {Kalman} {Filter}—{Part} {I}: {Fundamentals}},
	volume = {71},
	issn = {1941-0476},
	shorttitle = {Inverse {Extended} {Kalman} {Filter}—{Part} {I}},
	doi = {10.1109/TSP.2023.3304761},
	journal = {IEEE Transactions on Signal Processing},
	author = {Singh, Himali and Chattopadhyay, Arpan and Mishra, Kumar Vijay},
	year = {2023},
	pages = {2936--2951},
}

@misc{fang_inverse_2025,
	title = {The inverse {Kalman} filter},
	doi = {10.48550/arXiv.2407.10089},
	publisher = {arXiv},
	author = {Fang, Xinyi and Gu, Mengyang},
	month = jun,
	year = {2025},
	note = {arXiv:2407.10089 [stat]},
}

@misc{lunde_ensemble_2025,
	title = {An {Ensemble} {Information} {Filter}: {Retrieving} {Markov}-information from the {SPDE} discretisation},
	shorttitle = {An {Ensemble} {Information} {Filter}},
	doi = {10.48550/arXiv.2501.09016},
	publisher = {arXiv},
    author = {Lunde, Berent {\AA}nund Str{\o}mnes},
    month = jan,
	year = {2025},
	note = {arXiv:2501.09016 [stat]},
}

@article{gaspari_construction_1999,
	title = {Construction of correlation functions in two and three dimensions},
	volume = {125},
	copyright = {Copyright © 1999 Royal Meteorological Society},
	issn = {1477-870X},
	doi = {10.1002/qj.49712555417},
	language = {en},
	number = {554},
	journal = {Quarterly Journal of the Royal Meteorological Society},
	author = {Gaspari, Gregory and Cohn, Stephen E.},
	year = {1999},
	pages = {723--757},
}

@article{bishop_adaptive_2001,
	title = {Adaptive {Sampling} with the {Ensemble} {Transform} {Kalman} {Filter}. {Part} {I}: {Theoretical} {Aspects}},
	volume = {129},
	issn = {1520-0493, 0027-0644},
	shorttitle = {Adaptive {Sampling} with the {Ensemble} {Transform} {Kalman} {Filter}. {Part} {I}},
	doi = {10.1175/1520-0493(2001)129<0420:ASWTET>2.0.CO;2},
	language = {EN},
	number = {3},
	journal = {Monthly Weather Review},
	author = {Bishop, Craig H. and Etherton, Brian J. and Majumdar, Sharanya J.},
	month = mar,
	year = {2001},
	note = {Publisher: American Meteorological Society
Section: Monthly Weather Review},
	pages = {420--436},
}

@article{bannister_review_2017,
	title = {A review of operational methods of variational and ensemble-variational data assimilation},
	volume = {143},
	copyright = {© 2016 The Authors. Quarterly Journal of the Royal Meteorological Society published by John Wiley \& Sons Ltd on behalf of the Royal Meteorological Society.},
	issn = {1477-870X},
	doi = {10.1002/qj.2982},
	language = {en},
	number = {703},
	journal = {Quarterly Journal of the Royal Meteorological Society},
	author = {Bannister, R. N.},
	year = {2017},
	pages = {607--633},
}

@article{ott_local_2004,
	title = {A local ensemble {Kalman} filter for atmospheric data assimilation},
	volume = {56},
	issn = {null},
	doi = {10.3402/tellusa.v56i5.14462},
	number = {5},
	journal = {Tellus A: Dynamic Meteorology and Oceanography},
	author = {Ott, Edward and , Brian R., Hunt and , Istvan, Szunyogh and , Aleksey V., Zimin and , Eric J., Kostelich and , Matteo, Corazza and , Eugenia, Kalnay and , D.J., Patil and and Yorke, James A.},
	month = jan,
	year = {2004},
	note = {Publisher: Taylor \& Francis},
	pages = {415--428},
}

@article{hamill_distance-dependent_2001,
	title = {Distance-{Dependent} {Filtering} of {Background} {Error} {Covariance} {Estimates} in an {Ensemble} {Kalman} {Filter}},
	volume = {129},
	issn = {1520-0493, 0027-0644},
	doi = {10.1175/1520-0493(2001)129<2776:DDFOBE>2.0.CO;2},
	language = {EN},
	number = {11},
	journal = {Monthly Weather Review},
	author = {Hamill, Thomas M. and Whitaker, Jeffrey S. and Snyder, Chris},
	month = nov,
	year = {2001},
	note = {Publisher: American Meteorological Society
Section: Monthly Weather Review},
	pages = {2776--2790},
}

@article{bishop_ensemble_2009,
	title = {Ensemble covariances adaptively localized with {ECO}-{RAP}. {Part} 1: tests on simple error models},
	volume = {61},
	issn = {1600-0870},
	shorttitle = {Ensemble covariances adaptively localized with {ECO}-{RAP}. {Part} 1},
	language = {en-US},
	number = {1},
	journal = {Tellus A: Dynamic Meteorology and Oceanography},
	author = {Bishop, Craig H. and Hodyss, Daniel},
	month = jan,
	year = {2009},
}

@article{bishop_ensemble_2009-1,
	title = {Ensemble covariances adaptively localized with {ECO}-{RAP}. {Part} 2: a strategy for the atmosphere},
	volume = {61},
	issn = {1600-0870},
	shorttitle = {Ensemble covariances adaptively localized with {ECO}-{RAP}. {Part} 2},
	language = {en-US},
	number = {1},
	journal = {Tellus A: Dynamic Meteorology and Oceanography},
	author = {Bishop, Craig H. and Hodyss, Daniel},
	month = jan,
	year = {2009},
}

@article{bishop_adaptive_2011,
	title = {Adaptive {Ensemble} {Covariance} {Localization} in {Ensemble} {4D}-{VAR} {State} {Estimation}},
	volume = {139},
	issn = {1520-0493, 0027-0644},
	doi = {10.1175/2010MWR3403.1},
	language = {EN},
	number = {4},
	journal = {Monthly Weather Review},
	author = {Bishop, Craig H. and Hodyss, Daniel},
	month = apr,
	year = {2011},
	note = {Publisher: American Meteorological Society
Section: Monthly Weather Review},
	pages = {1241--1255},
}

@article{anderson_localization_2012,
	title = {Localization and {Sampling} {Error} {Correction} in {Ensemble} {Kalman} {Filter} {Data} {Assimilation}},
	volume = {140},
	issn = {1520-0493, 0027-0644},
	doi = {10.1175/MWR-D-11-00013.1},
	language = {EN},
	number = {7},
	journal = {Monthly Weather Review},
	author = {Anderson, Jeffrey L.},
	month = jul,
	year = {2012},
	note = {Publisher: American Meteorological Society
Section: Monthly Weather Review},
	pages = {2359--2371},
}

@article{vishny_high-dimensional_2024,
	title = {High-{Dimensional} {Covariance} {Estimation} {From} a {Small} {Number} of {Samples}},
	volume = {16},
	copyright = {© 2024 The Author(s). Journal of Advances in Modeling Earth Systems published by Wiley Periodicals LLC on behalf of American Geophysical Union.},
	issn = {1942-2466},
	language = {en},
	number = {9},
	journal = {Journal of Advances in Modeling Earth Systems},
	author = {Vishny, David and Morzfeld, Matthias and Gwirtz, Kyle and Bach, Eviatar and Dunbar, Oliver R. A. and Hodyss, Daniel},
	year = {2024},
	pages = {e2024MS004417},
}

@book{jazwinski_stochastic_1970,
	edition = {1st},
	title = {Stochastic {Processes} and {Filtering} {Theory}},
	volume = {64},
	author = {Jazwinski, Andrew H.},
	month = jan,
	year = {1970},
}

@article{eldering_orbiting_nodate,
	title = {The {Orbiting} {Carbon} {Observatory}-2: first 18 months of science data products},
	shorttitle = {The {Orbiting} {Carbon} {Observatory}-2},
	language = {en},
	author = {Eldering, Annmarie and O'Dell, Chris W. and Wennberg, Paul O. and Crisp, David and Gunson, Michael R. and Viatte, Camille and Avis, Charles and Braverman, Amy and Castano, Rebecca and Chang, Albert and Chapsky, Lars and Cheng, Cecilia and Connor, Brian and Dang, Lan and Doran, Gary and Fisher, Brendan and Frankenberg, Christian and Fu, Dejian and Granat, Robert and Hobbs, Jonathan and Lee, Richard A. M. and Mandrake, Lukas and McDuffie, James and Miller, Charles E. and Myers, Vicky and Natraj, Vijay and O'Brien, Denis and Osterman, Gregory B. and Oyafuso, Fabiano and Payne, Vivienne H. and Pollock, Harold R. and Polonsky, Igor and Roehl, Coleen M. and Rosenberg, Robert and Schwandner, Florian and Smyth, Mike and Tang, Vivian and Taylor, Thomas E. and To, Cathy and Wunch, Debra and Yoshimizu, Jan},
}

@misc{brodrick_shift_2024,
	title = {{SHIFT}: {AVIRIS}-{NG} {L2A} {Unrectified} {Surface} {Reflectance} {Version} 1},
	language = {en},
	publisher = {ORNL Distributed Active Archive Center},
	author = {Brodrick, P.G. and Pavlick, R. and Bernas, M. and Chapman, J.W. and Eckert, R. and Helmlinger, M. and Hess-Flores, M. and Rios, L.M. and Schneider, F.D. and Smyth, M.M. and Eastwood, M. and Green, R.O. and Thompson, D.R. and Chadwick, K.D. and Schimel, D.S.},
	year = {2024},
	doi = {10.3334/ORNLDAAC/2376},
}

@article{ramgraber_ensemble_2023-1,
	title = {Ensemble transport smoothing. {Part} {II}: {Nonlinear} updates},
	volume = {17},
	issn = {2590-0552},
	shorttitle = {Ensemble transport smoothing. {Part} {II}},
	doi = {10.1016/j.jcpx.2023.100133},
	journal = {Journal of Computational Physics: X},
	author = {Ramgraber, Maximilian and Baptista, Ricardo and McLaughlin, Dennis and Marzouk, Youssef},
	month = nov,
	year = {2023},
	pages = {100133},
}

@book{sarkka_bayesian_2013,
	series = {Institute of {Mathematical} {Statistics} {Textbooks}},
	title = {Bayesian {Filtering} and {Smoothing}},
	publisher = {Cambridge University Press},
	author = {Särkkä, Simo},
	year = {2013},
}

@article{chadwick_unlocking_2025,
	title = {Unlocking ecological insights from sub-seasonal visible-to-shortwave infrared imaging spectroscopy: {The} {SHIFT} campaign},
	volume = {16},
	copyright = {© 2025 The Author(s). Ecosphere published by Wiley Periodicals LLC on behalf of The Ecological Society of America.},
	issn = {2150-8925},
	shorttitle = {Unlocking ecological insights from sub-seasonal visible-to-shortwave infrared imaging spectroscopy},
	doi = {10.1002/ecs2.70194},
	language = {en},
	number = {3},
	journal = {Ecosphere},
	author = {Chadwick, K. Dana and Davis, Frank and Miner, Kimberley R. and Pavlick, Ryan and Reynolds, Mark and Townsend, Philip A. and Brodrick, Philip G. and Ade, Christiana and Allen, Jean and Anderegg, Leander and Angel, Yoseline and Boving, Indra and Byrd, Kristin B. and Campbell, Petya and Carberry, Luke and Cavanaugh, Katherine C. and Cavanaugh, Kyle C. and Easterday, Kelly and Eckert, Regina and Gierach, Michelle and Gold, Kaitlin and Hestir, Erin and Huemmrich, Fred and Klope, Maggie and Kokaly, Raymond F and Lovegreen, Piper and Luis, Kelly and McMahon, Conor and Nidzieko, Nicholas and Ochoa, Francisco and Ongjoco, Anna Jiselle and Ordway, Elsa and Pascolini-Campbell, Madeleine and Queally, Natalie and Roberts, Dar A. and Saiki, Clare M. and Schneider, Fabian D. and Shiklomanov, Alexey N. and Silva, Germán D. and Snyder, Jordan and Thornton, Michele and Trugman, Anna T. and Vinod, Nidhi and Zheng, Ting and Avouris, Dulcinea M. and Baker, Brianna and Baskaran, Latha and Bell, Tom and Berg, Megan and Bernas, Michael and Bohn, Niklas and Braghiere, Renato K. and Breuer, Zach and Brooks, Andrew J. and Burkard, Nolan and Burmistrova, Julia and Cawse-Nicholson, Kerry and Chapman, John and Chazaro-Haraksin, Johana and Cryer, Joel and Cushman, K. C. and Dahlin, Kyla and Dao, Phuong D. and DiBartolo, Athena and Eastwood, Michael and Elder, Clayton D. and Giordani, Angela and Grant, Kathleen and Green, Robert O. and Hanson, Alexa and Heberlein, Brendan and Helmlinger, Mark and Hook, Simon and Jensen, Daniel and Johnson, Emma and Johnson, Marie and Kiper, Michael and Kibler, Christopher and King, Jennifer Y. and Kovach, Kyle R. and Kreisberg, Aaron and Lacey, Daniel and Lang, Evan and Lee, Christine and Lopez, Amanda M. and Lopez Barreto, Brittany and Maguire, Andrew and Marsh, Elliott and Miller, Charles and Nguyen, Dieu My T. and Nickles, Cassandra and Ocón, Jonathan P. and Papen, Elijah P. and Park, Maria and Poulter, Benjamin and Raiho, Ann and Reim, Porter and Robinson, Timothy H. and Romero Galvan, Fernando E. and Shafron, Ethan and Skalitzky, Brenen R. and Stroschein, Sydney and Taylor, Niky C. and Thompson, David R. and Thompson, Kate and Tye, Cecily and Van Beek, Joelie and Vanden Heuvel, Cecilia and Vellanoweth, Jonathan and Vermeer, Evert and Villanueva-Weeks, Claire and Zumdahl, Kristen and Schimel, David},
	year = {2025},
	pages = {e70194},
}

@article{thompson_-orbit_2024,
	title = {On-orbit calibration and performance of the {EMIT} imaging spectrometer},
	volume = {303},
	issn = {0034-4257},
	doi = {10.1016/j.rse.2023.113986},
	journal = {Remote Sensing of Environment},
	author = {Thompson, David R. and Green, Robert O. and Bradley, Christine and Brodrick, Philip G. and Mahowald, Natalie and Dor, Eyal Ben and Bennett, Matthew and Bernas, Michael and Carmon, Nimrod and Chadwick, K. Dana and Clark, Roger N. and Coleman, Red Willow and Cox, Evan and Diaz, Ernesto and Eastwood, Michael L. and Eckert, Regina and Ehlmann, Bethany L. and Ginoux, Paul and Ageitos, María Gonçalves and Grant, Kathleen and Guanter, Luis and Pearlshtien, Daniela Heller and Helmlinger, Mark and Herzog, Harrison and Hoefen, Todd and Huang, Yue and Keebler, Abigail and Kalashnikova, Olga and Keymeulen, Didier and Kokaly, Raymond and Klose, Martina and Li, Longlei and Lundeen, Sarah R. and Meyer, John and Middleton, Elizabeth and Miller, Ron L. and Mouroulis, Pantazis and Oaida, Bogdan and Obiso, Vincenzo and Ochoa, Francisco and Olson-Duvall, Winston and Okin, Gregory S. and Painter, Thomas H. and Pérez García-Pando, Carlos and Pollock, Randy and Realmuto, Vincent and Shaw, Lucas and Sullivan, Peter and Swayze, Gregg and Thingvold, Erik and Thorpe, Andrew K. and Vannan, Suresh and Villarreal, Catalina and Ung, Charlene and Wilson, Daniel W. and Zandbergen, Sander},
	month = mar,
	year = {2024},
	pages = {113986},
}

@article{meister_ocean_2024,
	title = {The Ocean Color Instrument ({OCI}) on the Plankton, Aerosol, Cloud, ocean Ecosystem ({PACE}) Mission: System Design and Prelaunch Radiometric Performance},
	volume = {62},
	issn = {1558-0644},
	doi = {10.1109/TGRS.2024.3383812},
	shorttitle = {The Ocean Color Instrument ({OCI}) on the Plankton, Aerosol, Cloud, ocean Ecosystem ({PACE}) Mission},
	pages = {1--18},
	journaltitle = {{IEEE} Transactions on Geoscience and Remote Sensing},
	author = {Meister, Gerhard and Knuble, Joseph J. and Gliese, Ulrik and Bousquet, Robert and Chemerys, Leland H. and Choi, Hyeungu and Eplee, Robert E. and Estep, Robert H. and Gorman, Eric T. and Kitchen-{McKinley}, Samuel and Kubalak, David and Lee, Shihyan and {McClain}, Charles and {McIntire}, Jeffrey W. and Patt, Frederick S. and Rhodes, Zakk and Werdell, P. Jeremy},
	date = {2024},
}

@article{guanter_enmap_2015,
	title = {The {EnMAP} Spaceborne Imaging Spectroscopy Mission for Earth Observation},
	volume = {7},
	rights = {http://creativecommons.org/licenses/by/3.0/},
	issn = {2072-4292},
	doi = {10.3390/rs70708830},
	pages = {8830--8857},
	number = {7},
	journaltitle = {Remote Sensing},
	publisher = {Multidisciplinary Digital Publishing Institute},
	author = {Guanter, Luis and Kaufmann, Hermann and Segl, Karl and Foerster, Saskia and Rogass, Christian and Chabrillat, Sabine and Kuester, Theres and Hollstein, André and Rossner, Godela and Chlebek, Christian and Straif, Christoph and Fischer, Sebastian and Schrader, Stefanie and Storch, Tobias and Heiden, Uta and Mueller, Andreas and Bachmann, Martin and Mühle, Helmut and Müller, Rupert and Habermeyer, Martin and Ohndorf, Andreas and Hill, Joachim and Buddenbaum, Henning and Hostert, Patrick and Van der Linden, Sebastian and Leitão, Pedro J. and Rabe, Andreas and Doerffer, Roland and Krasemann, Hajo and Xi, Hongyan and Mauser, Wolfram and Hank, Tobias and Locherer, Matthias and Rast, Michael and Staenz, Karl and Sang, Bernhard},
	date = {2015-07},
	langid = {english},
}

@article{krutz_instrument_2019,
	title = {The Instrument Design of the {DLR} Earth Sensing Imaging Spectrometer ({DESIS})},
	volume = {19},
	rights = {http://creativecommons.org/licenses/by/3.0/},
	issn = {1424-8220},
	doi = {10.3390/s19071622},
	pages = {1622},
	number = {7},
	journaltitle = {Sensors},
	publisher = {Multidisciplinary Digital Publishing Institute},
	author = {Krutz, David and Müller, Rupert and Knodt, Uwe and Günther, Burghardt and Walter, Ingo and Sebastian, Ilse and Säuberlich, Thomas and Reulke, Ralf and Carmona, Emiliano and Eckardt, Andreas and Venus, Holger and Fischer, Christian and Zender, Bernd and Arloth, Simone and Lieder, Matthias and Neidhardt, Michael and Grote, Ute and Schrandt, Friedrich and Gelmi, Samuele and Wojtkowiak, Andreas},
	date = {2019-01},
	langid = {english},
}

@article{cogliati_prisma_2021,
	title = {The {PRISMA} imaging spectroscopy mission: overview and first performance analysis},
	volume = {262},
	issn = {0034-4257},
	doi = {10.1016/j.rse.2021.112499},
	shorttitle = {The {PRISMA} imaging spectroscopy mission},
	pages = {112499},
	journaltitle = {Remote Sensing of Environment},
	shortjournal = {Remote Sensing of Environment},
	author = {Cogliati, S. and Sarti, F. and Chiarantini, L. and Cosi, M. and Lorusso, R. and Lopinto, E. and Miglietta, F. and Genesio, L. and Guanter, L. and Damm, A. and Pérez-López, S. and Scheffler, D. and Tagliabue, G. and Panigada, C. and Rascher, U. and Dowling, T. P. F. and Giardino, C. and Colombo, R.},
	date = {2021-09-01},
}

@book{nieke_towards_2018,
	title = {Towards the Copernicus Hyperspectral Imaging Mission For The Environment ({CHIME})},
	doi = {10.1109/IGARSS.2018.8518384},
	pagetotal = {157},
	author = {Nieke, Jens and Rast, Michael},
	date = {2018-07-01},
	note = {Pages: 159},
}

\end{document}